\documentstyle[12pt]{article}
\newcommand{\h}{\hspace*{0.5 cm}}
\newcommand{\ppar}{{\vspace*{18truept}\par}}
\newcommand{\parn}{\par\noindent}
\newcommand{\sspace}{{\hbox{\hspace*{10 truept}}}}

\newcommand{\Rerm}{{\rm Re}}
\newcommand{\Imrm}{{\rm Im}}
\newcommand{\Rrm}{{\rm R}}
\newcommand{\Trm}{{\rm T}}
\newcommand{\Drm}{{\rm D}}
\newcommand{\Crm}{{\rm C}}
\newcommand{\Brm}{{\rm B}}
\newcommand{\Lrm}{{\rm L}}
\newcommand{\Srm}{{\rm S}}
\newcommand{\Irm}{{\rm I}}
\newcommand{\IIrm}{{\rm II}}
\newcommand{\IIIrm}{{\rm III}}
\newcommand{\grm}{{\rm g}}
\newcommand{\Inrm}{{\rm in}}
\newcommand{\BLrm}{{\rm BL}}
\newcommand{\inrm}{{\rm in}}
\newcommand{\pirm}{{\rm p}}
\newcommand{\frm}{{\rm f}}
\newcommand{\irm}{{\rm i}}
\newcommand{\clrm}{{\rm cl}}
\newcommand{\path}{{\rm path}}

 1
\font\ner=cmbx10
 1  
 2  
 1  
 
 1
 1
  
 1

\newcommand{\xx}{{$\times$}}
\newcommand{\cof}{{\em cut-off}}
\newcommand{\coff}{{\em cut-off}}
\newcommand{\drm}{{\rm d}}

\newcommand{\pht}{{\em phase time}}
\newcommand{\tdl}{{\em tempi di Larmor}}
\newcommand{\dwt}{{\em dwell time}}
\newcommand{\tes}{{\em tempo di fase estrapolato}}
\newcommand{\tess}{{\em tempi di fase estrapolati}}
\newcommand{\teq}{{\em tempo equivalente}}
\newcommand{\tbl}{{\em tempo di B\"uttiker-Landauer}}

\newcommand{\phtz}{{\em phase time }}
\newcommand{\tdlz}{{\em tempi di Larmor }}
\newcommand{\dwtz}{{\em dwell time }}

\newcommand{\teqz}{{\em tempo equivalente }}

\newcommand{\Dwtz}{{\em Dwell time }}

\newcommand{\tpen}{{$\overline{\tau_{{\rm Pen}}}(0,x)$ }}
\newcommand{\ttun}{{$\overline{\tau_{{\rm Tun}}}(0,d)$ }}
\newcommand{\tret}{{$\overline{\tau_{{\rm Ret}}}(x,x)$ }}

\begin{document}

\centerline{\large\bf Tempi di Tunnelling (Tunneling Times)$^{(*)}$}
\footnotetext{$^{(*)}$ Lavoro in parte finanziato da INFN, MURST, CNR 
(Italia), da CAPES, CNPq (Brasile), e dall' I.N.R. (Accademia Ucraina
delle Scienze, Kiev).}
 
\vspace*{1 cm}
 
\centerline{Giuseppe PRIVITERA}

\vspace*{0.5 cm}
 
{\small
\centerline{{\em Dipartimento di Fisica, Universit\`a
di Catania, Catania, Italia}}

\vspace*{1. cm}

\centerline{Erasmo RECAMI and  Vladislav S. OLKHOVSKY$^{(a)}$}
 
\vspace*{0.5 cm}
 
{\small
\centerline{{\em Facolt\`a di Ingegneria, Universit\`a
statale di Bergamo, Bergamo, Italia;}}
\centerline{\em I.N.F.N., Sezione di Milano, Milano, Italia; and}
\centerline{\em DMO/FEEC and C.C.S., Universit\`a  Statale di
Campinas, S.P., Brasile.}
\centerline{$^{(a)}$ {\em Institute for Nuclear Research, Ukrainian
Academy of Sciences, Kiev.}}
}
 
\vspace*{1. cm}
 
{\bf ABSTRACT --} \ In this paper (in Italian) we critically and detaily
examine
various definitions existing in the literature for the tunnelling times:
namely, the phase-time; the centroid-based times; the B\"uttiker and Landauer 
times; the Larmor times; the complex (path-integral and Bohm) times; the
dwell time, and the generalized (Olkhovsky and Recami) dwell time, with some
numerical evaluations. \ Then, we pass to examine the equivalence between
quantum tunnelling and ``photon tunnelling" (evanescent waves propagation),
with particular attention to tunnellings with Superluminal group-velocities
(``Hartman effect"). \ At last, the main (Cologne, Berkeley, Florence,
Vienna) experimental data about Superluminal evanescent wave propagation
are briefly reviewed.\\

\
\centerline{{\bf Parte I:  TEORIA}}

\

\

{\bf 
I.0) Introduzione.}\\ \rm

Consideriamo una particella inizialmente libera che,  
durante il  proprio moto, vada ad incidere
su una barriera di potenziale di energia maggiore della propria.
\par
Come sappiamo la meccanica quantistica prevede che vi sia  una  probabilit\`a 
non nulla che la particella possa attraversare la barriera
({\em effetto tunnel\/}).
\`E dunque lecito chiedersi se sia possibile associare un tempo, e quindi una 
velocit\`a, al processo di attraversamento della barriera e, in tal caso, 
come sia possibile calcolare e misurare sperimentalmente queste quantit\`a.\par
Potrebbe sorprendere che una domanda apparentemente cos\'{\i} semplice 
e di fondamentale importanza nella comprensione di qualunque teoria 
dello scattering, e non solo, non abbia ancora ricevuto una adeguata risposta.
\ppar 
Il problema venne posto per la prima volta nel 1931 da Condon,$[1]$ 
ed un primo tentativo di risposta risale all'anno successivo da parte
di McColl.$[2]$ Da allora, per\`o, l'argomento rimase quasi ignorato 
fino agli anni quaranta e cinquanta, con il tentativo
di introdurre una osservabile quantomeccanica per la variabile tempo, 
soprattutto nella trattazione delle collisioni [per una semplice introduzione
in M.Q. di un operatore (non autoaggiunto ma) hermitiano per l'osservabile 
Tempo, si vedano le refs.$[3]$.]\par 
L'avvento di dispositivi elettronici ad alta velocit\`a, basati su processi 
di tunnelling, e l'importanza del fenomeno in processi di fissione e 
fusione nucleare sotto soglia, hanno per\`o suscitato, soprattutto negli ultimi 
quindici anni, un crescente interesse verso l'argomento, tanto da portare,
dal 1987 ad oggi, alla pubblicazione di parecchie 
(almeno una decina$[4-6]$) reviews teoriche al riguardo.\ppar 
Sperimentalmente, invece, le difficolt\`a nell'effettuare misure non 
invasive con particelle quali ad esempio elettroni, hanno reso molto difficile 
una verifica dei risultati teorici. 
Solo negli ultimi anni, sfruttando l'equivalenza tra trasmissione di
onde elettromagnetiche 
evanescenti$[7]$ e tunnelling di particelle,
 sono state realizzate alcune misure dei tempi di
trasmissione di microonde e di fotoni. Ci occuperemo in seguito, quando 
prenderemo in esame tali esperimenti, della suddetta equivalenza.\ppar 
Poich\'e effetti di tunnelling possono intervenire in vari processi 
fisici (I.teoria dello scattering, diseccitazione di stati metastabili, 
fissione e fusione sottosoglia, etc.) premettiamo che noi esamineremo 
solo il caso unidimensionale di attraversamento, 
da parte di una particella inizialmente libera, di una barriera 
di potenziale costante$^{\#1}$ $V_0$,
\footnotetext{$^{\#1}$ Solo in un caso, quello del  \tbl,
la barriera  sar\`a anche funzione del tempo.} 
che si estenda nell'intervallo $[0,d]$, (vedi Fig.I-1); caso peraltro
equivalente alle configurazioni degli apparati sperimentali finora adottati.

\h Malgrado la gran  mole di lavori teorici cui accennavamo, non esiste 
ancora un approccio al problema che venga universalmente accettato; noi li 
raggrupperemo in quattro categorie principali.\ppar
Alla prima categoria appartengono tutti quei tempi costruiti ``seguendo" il 
pacchetto incidente durante l'attraversamento della barriera.\\ 

\

Figura I-1.\\

\

\h Scelta 
una data caratteristica del pacchetto, per esempio il picco centrale, si 
confrontano il picco entrante e quello uscente in modo da ricavarne una 
relazione temporale. Oltre al picco sono stati presi in considerazione  
dai vari autori anche il centroide (centro di ``massa" del pacchetto) 
e il fronte d'onda, molto netto, di un'onda a scalino.\ppar
Tra le critiche pi\'u comunemente avanzate verso questo tipo di 
approccio, vi \`e il fatto che, secondo molti, il picco 
uscente non  corrisponderebbe sempre a quello entrante
a causa della presenza,
tra le componenti di Fourier del pacchetto, di frequenze corrispondenti 
ad energie prossime o al di sopra dell'energia di barriera. 
Tali frequenze raggiungerebbero 
prima delle altre la barriera venendo da questa trasmesse pi\'u 
efficientemente di quelle pi\'u basse. Tutto ci\`o, oltre a provocare 
gli usuali effetti di dispersione, eventualmente presenti anche durante la 
fase di avvicinamento alla barriera, sembra 
poter avere, in certi casi, un effetto accelerante sul pacchetto o, 
addirittura,
portare alla sua trasmissione prima ancora che il picco principale la  
abbia raggiunta.\ppar
A questa categoria di tempi appartiene il \pht . Questo viene ricavato mediante
l'approssimazione della fase stazionaria utilizzando la definizione di 
velocit\`a di gruppo del pacchetto e, malgrado le suddette critiche, risulta 
quello maggiormente in accordo con i dati sperimentali finora a disposizione.
\ppar
Un secondo tipo di approccio consiste nell'introduzione di qualche grado di 
libert\`a nel sistema particella-barriera, per definire un
``orologio" che dia una misura del tempo trascorso dalla particella 
all'interno della barriera.
Attraverso l'effetto dell'orologio sulla particella o, viceversa, attraverso
l'effetto della barriera su un orologio associato alla particella, 
si risale al loro tempo di interazione durante l'attraversamento.\par
B\"uttiker e Landauer, ad esempio, tentano di 
risalire al tempo di tunnelling calcolando lo scambio di quanti di energia da 
parte di una particella che attraversi una barriera rettangolare la cui
altezza sia modulata nel tempo  (\tbl ).\par
Un altro esempio pu\`o essere la misura della rotazione dello spin di un 
elettrone quando nella porzione di spazio occupata dalla barriera si trovi un 
campo magnetico uniforme (\tdl ).\ppar 
Questo tipo di approccio consente una larga scelta nei gradi di libert\`a del 
sistema che possono essere usati come orologio e, contemporaneamente, 
fornisce anche dei metodi sperimentali per la verifica dei risultati teorici.
Viene per\`o criticato da alcuni autori, sia perch\'e non tutti i tipi di 
orologio risultano equivalenti, sia soprattutto perch\'e l'introduzione di
tali tipi di orologi, oltre a modificare il numero di gradi di
libert\`a del sistema, introduce comunque dei processi 
invasivi, i quali condizionerebbero i risultati. Praticamente niente ci
assicura che il tempo di interazione, in cui comunque lo stato del sistema 
subisce una perturbazione, corrisponda proprio al tempo di attraversamento o 
di riflessione.\ppar
Il terzo tipo di approccio al problema  consiste nell' attribuire al moto
della particella sotto barriera un insieme di traiettorie ``semiclassiche".
Con queste viene poi calcolato un tempo medio di tunnelling.\par
I cammini possono essere  costruiti con vari metodi, per esempio attraverso 
i {\em path-integrals} di Feynman, il metodo di Bohm, o l'uso della 
distribuzione 
di Wigner: tutti e tre i metodi portano, naturalmente ad una distribuzione di 
tempi.\par
Uno degli inconvenienti di questo tipo di approccio \`e quello di fornire 
dei risultati complessi. Tuttavia \`e possibile estrarre
da questi delle quantit\`a reali (il modulo,
la parte reale  e la parte immaginaria) che, in certi casi,
risultano essere strettamente 
connesse con i tempi definiti mediante altri approcci. Proprio
per questo motivo l'interpretazione fisica di tali risultati appare
comunque molto interessante.
\ppar
L'ultima categoria di tempi di tunnelling che esamineremo \`e quella che parte 
dalla definizione di \dwtz o tempo di soggiorno. Quest'ultimo 
viene definito come: 
$$\tau^\Drm(x_1,x_2;k)=j^{-1}\int_{x_1}^{x_2}|\psi (x,k)|^2 \drm x, \eqno(I.0.0)$$
e rappresenta il tempo 
speso da una particella all'interno della barriera, o di una qualunque 
regione di spazio, calcolato come rapporto tra densit\`a di probabilit\`a che 
la particella si trovi in quella regione ed il flusso $j_{\inrm}$ in essa 
entrante.
\ppar
Il problema di tale definizione \`e che questa ci fornisce si 
il tempo di soggiorno all'interno 
della barriera, ma senza distinguere tra canale di trasmissione e canale di 
riflessione. A tal proposito, una relazione, spesso  considerata ovvia ma non 
universalmente accettata, 
che dovrebbe legare i tempi corrispondenti ai due canali \`e: 
$$\tau_\Drm=|T(k)|^2\tau_\Trm + |R(k)|^2\tau_\Rrm. \eqno(I.0.1)$$ 
Questa relazione, se pure fosse  corretta,
sembra non bastare da sola ad individuare univocamente $\tau_\Trm$ e $\tau_\Rrm$.\par
Le critiche a tale relazione, ed i tentativi di alcuni autori di trovare una 
relazione equivalente che leghi il tempo trascorso all'interno della barriera 
con i tempi di riflessione e di attraversamento, portano ai cosiddetti 
approcci spaziali. Questi vengono costruiti attraverso l'interpretazione 
probabilistica 
standard, in meccanica quantistica, delle densit\`a di corrente relative a
trasmissione e riflessione. 
\ppar
Notiamo infine che alcuni dei tempi che descriveremo, potrebbero sembrare 
dei candidati migliori di altri a rappresentare il tempo di tunnelling. Ci\`o 
sia perch\'e non presentano il comparire dell'{\em effetto Hartman}, sia 
perch\'e in accordo con la (I.0.1). Tuttavia proprio questi tempi risultano, 
invece, maggiormente in disaccordo con i dati sperimentali.\\ 

\

{\bf 
I.1) Premesse e notazioni.}\\ \rm

Prima di andare a descrivere i pi\'u importanti tipi di tempi ed i relativi 
risultati premettiamo alcune considerazioni e notazioni.\parn
Supporremo di aver gi\`a risolto in modo esatto il caso stazionario. Per 
ogni energia fissata, $E=\hbar ^2 k^2/2m$, sar\`a allora:
$$\psi (x;k)=\cases{\psi_\Irm= e^{ikx}+R(k)e^{-i(kx-\beta)} & $x\leq 0$\cr
		  \psi_{\IIrm}=\chi (x;k) & $0\leq x\leq d$\cr
		  \psi_{\IIIrm}=T(k)e^{i(kx+\alpha)} &$ x\geq d$\cr}\eqno(I.1.0)$$
con $R(k)$ e $T(k)$ ampiezze, rispettivamente, di riflessione e di 
trasmissione, 
tali che: $$R(k)=\sqrt{1-T(k)^2},$$ 
e $\beta =\beta (k)$, $\alpha =\alpha (k)$, corrispondenti ritardi di fase.
\parn
Nel caso particolare di una barriera rettangolare:
$$V(x)=\cases{V_0 & $0\leq x\leq d$\cr
		    ${\rm 0}$ & {\rm altrove},\cr }$$
$\chi (x;k),\;R(k),\;T(k),\;\alpha (k),\;\beta (k)\;$, sono tutte quantit\`a 
la cui forma analitica \`e ben conosciuta; avremo dunque:
$$\chi (x;k)=\cases{A(k)e^{-\kappa x}+B(k)e^{\kappa x} & $E < V_0$\cr
A(k)e^{-i\kappa x}+B(k)e^{i\kappa x} & $E > V_0$\cr},\eqno(I.1.1)$$\noindent 
con $$\kappa =\cases{\sqrt{2m(V_0-E)}/\hbar,&$E < V_0$\cr
			     \sqrt{2m(E-V_0)}/\hbar,&$E > V_0$\cr}$$
$$ R(k)=\cases{\displaystyle{(k^2-\kappa^2)\sinh (\kappa d)\over
[4k^2\kappa^2+(k^2+\kappa^2)^2\sinh^2(\kappa d)]^{1\over2}}&
$E < V_0$\cr\displaystyle {(k^2+\kappa^2)\sin (\kappa d)\over
[4k^2\kappa^2+(k^2-\kappa^2)^2\sin^2(\kappa d)]^{1\over2}}&
$E > V_0$\cr}\eqno(I.1.2)$$
$$ T(k)=\cases{\displaystyle{2k\kappa\over
[4k^2\kappa^2+(k^2+\kappa^2)^2\sinh^2(\kappa d)]^{1\over2}}&
$E < V_0$\cr\displaystyle {2k\kappa\over
[4k^2\kappa^2 +(k^2-\kappa^2)^2\sin^2(\kappa d)]^{1\over2}}&
$E > V_0$\cr}\eqno(I.1.3)$$
$$\beta (k)=\cases{\arctan\left(\displaystyle{-2k\kappa\over(k^2-\kappa^2)} 
\coth (\kappa d)\right),&$E < V_0$\cr
\arctan\left(\displaystyle{-2k\kappa\over(k^2+\kappa^2)} 
\cot (\kappa d)\right), &$E > V_0$\cr}\eqno(I.1.4)$$
$$\alpha (k)=\cases{\arctan\left(\displaystyle{(k^2-\kappa^2)\over 2k\kappa} 
\tanh (\kappa d)\right),&$E < V_0$\cr
\arctan\left(\displaystyle{(k^2+\kappa^2)\over 2k\kappa} 
\tan (\kappa d)\right), &$E > V_0$\cr}\eqno(I.1.5)$$

Naturalmente in seguito non avremo a che fare con semplici onde stazionarie 
ma con pacchetti d'onda della forma:
$$\Psi (x;t)=\int \drm k\ C f(k-k_0)\psi (x;k)e^{-i{E(k)t\over\hbar}}=$$
$$=\int \drm E\  g(E-E_0)\psi (x;k(E))e^{-i{Et\over\hbar}}.\eqno(I.1.6)$$
Saranno inoltre: 
$$\rho =|\psi (x)|^2,\sspace j=\Rerm \ \left\{ {i\hbar\over 2m}\psi(x)
{\partial\over\partial x}\psi^* (x)\right\} .\eqno(I.1.7)$$     
Nel caso elettromagnetico (microonde e fotoni), che noi supporremo sempre
T.E. o T.M., $\psi$ rappresenter\`a la parte scalare di 
una delle due conponenti del campo.
\ppar
Definiamo infine \teq, 
$\tau_{\rm eq}^\Trm$, il tempo che la particella impiegherebbe ad 
attraversare lo spazio occupato dalla barriera, ma in sua assenza; 
sar\`a dunque $\tau^\Trm_{\rm eq}=md/\hbar k$. 
Prendiamo ora in rassegna le definizioni di alcuni dei tempi cui abbiamo 
accennato.\\ 

\

{\bf 
I.2) Phase Time.}\\ \rm

Supponiamo di avere un pacchetto d'onda molto stretto intorno ad un numero 
d'onda $k_0$. La descrizione della sua evoluzione temporale \`e in genere 
molto complicata a causa, spesso, della propria natura dispersiva.
In ogni caso, sotto opportune condizioni, \`e possibile seguire 
la posizione del picco di 
un pacchetto simmetrico con buona precisione, trascurando tali effetti$[8]$.
Possiamo quindi provare ad identificare il pacchetto prendendo come 
riferimento il picco: per far ci\`o usiamo il metodo della fase 
stazionaria.\par 
Il picco del pacchetto sar\`a formato da quelle componenti di Fourier per  cui 
la variazione di fase in un intorno di $k_0$ sia abbastanza ridotta in modo 
da non interferire distruttivamente. 
Analogamente il pacchetto trasmesso sar\`a
descritto da una funzione d'onda anch'essa dominata da una piccola serie di 
frequenze ciascuna della forma:
$$\psi (x;k)\sim e^{i(kx-{E(k)t\over\hbar}+\alpha (k))}.$$ 
Se vogliamo seguire la posizione del picco, dobbiamo vedere per
quali valori di $x_\pirm(t)$ la fase \`e stazionaria, cio\`e, l'argomento 
dell'esponenziale \`e massimo.$^{\#2}$
\footnotetext{$^{\#2}$ Applicando lo stesso ragionamento alla parte incidente
otterremmo: $x_\pirm(t)={1\over\hbar}\left(
{\drm E\over \drm k}\right) t=\left({\drm \omega\over \drm k}
\right)t$, dove $(\drm  \omega/\drm k)$ \`e proprio, per sua definizione, la velocit\`a 
di gruppo del 
pacchetto incidente. Notiamo inoltre che, naturalmente, $v_\grm=(\drm  \omega/\drm k)$ 
dipender\`a dalla relazione di 
dispersione $\omega(k)$ del mezzo in cui il pacchetto si propaga.}.
Deve quindi essere:
$${\drm \over \drm k}\left(kx_\pirm(t)-{E(k)t\over\hbar}+\alpha (k)\right)=0$$
$$\Rightarrow \sspace x_\pirm(t)={1\over\hbar}{\drm E\over \drm k}t-{\drm \alpha\over \drm k}
\eqno(I.2.0)$$
Allora $\alpha^\prime (k_0)=(\drm \alpha/\drm k)_{k_0}$ rappresenter\`a il ritardo 
spaziale $\delta x$ causato dal processo di tunnelling e, dividendo per $v_\grm$
(velocit\`a di gruppo del pacchetto d'onda) otteniamo 
un ritardo temporale:
$$\delta\tau_\Trm={\delta x\over v_\grm}=(v_\grm)^{-1}{\drm \alpha\over \drm k}
=\left({1\over\hbar }{\drm E\over \drm k}\right)^{-1}{\drm \alpha\over \drm k}
=\hbar {\drm \alpha\over \drm E}.\eqno(I.2.1)$$ \\

\

Figura I-2.\\

\

\h Notiamo che stiamo calcolando sia $v_\grm$ che tutte le altre 
quantit\`a in $k=k_0$. Vedremo pi\'u avanti se e quando  ci\`o possa considerarsi
corretto.
\ppar
Per motivi che spiegheremo tra poco definiamo \phtz il tempo totale
$\tau_\Trm^\varphi (x_1,x_2;k)$, speso da una particella tra due punti,
$x_1$ e $x_2$, esterni alla barriera e abbastanza lontani da essa, 
cio\`e: $x_1\ll 0$ e $x_2\gg d$. Avremo allora:
$$\tau_\Trm^\varphi (x_1,x_2;k)={1\over v_\grm}(x_2-x_1+\alpha^\prime (k))\eqno
(I.2.2)$$
ed applicando lo stesso tipo di ragionamento alle particelle 
riflesse:
$$\tau_\Rrm^\varphi (x_1,x_2;k)={1\over v_\grm}(-2x_1+\beta^\prime (k))\eqno(I.2.3)$$
\par
Poich\'e sia $\tau_\Trm^\varphi$ che $\tau_\Rrm^\varphi$ dipendono linearmente da
$x_1$ e $x_2$, potremmo pensare di estrapolare i tempi di attraversamento e  
di riflessione direttamente facendo tendere $x_1$ e $x_2$ rispettivamente a
$0$ e $d$: ci\`o per\`o non \`e corretto. \par
Avendo infatti supposto le componenti del pacchetto strettamente
distribuite intorno ad un numero d'onda $k_0$, se $\Delta k=\sigma$,  
nello spazio ordinario avremo un pacchetto la cui larghezza sar\`a dell'ordine
di $\sigma^{-1}$ e, quindi, tanto pi\'u esteso quanto pi\'u esso sar\`a 
piccato intorno a $k_0$. Dunque la parte incidente e quella riflessa della 
funzione d'onda potranno interferire tra loro anche ad una certa 
distanza ($\sim\sigma^{-1}$) dalla barriera (vedi Fig.I-2).
Del resto, visto che stiamo usando un'approssimazione stazionaria, 
non stiamo seguendo realmente il pacchetto nella sua evoluzione temporale ma,
semplicemente, osservando asintoticamente il ritardo di fase corrispondente 
al numero d'onda $k_0$.\ppar
Definiamo comunque ugualmente:
$$\Delta\tau_\Trm^\varphi ={1\over v_\grm}(d+\alpha^\prime (k)),\eqno(I.2.4)$$
$$\Delta\tau_\Rrm^\varphi ={1\over v_\grm}(\beta^\prime (k)),\eqno(I.2.5)$$
che chiameremo \tess. In seguito, per\`o,
dobbiamo tenere sempre presente il carattere puramente asintotico di 
tali definizioni.\ppar 
Nel caso di barriera rettangolare abbiamo: 
$$\Delta\tau_\Trm^\varphi (0,d;k)=\Delta\tau_\Rrm^\varphi(0,d;k)=$$
$$={m\over\hbar k\kappa}\left({{2\kappa d k^2(\kappa^2-k^2)+\varepsilon ^4\sinh 
(2\kappa d)} \over{4\kappa ^2 k^2 + \varepsilon ^4\sinh^2(\kappa d)}}\right),
\eqno(I.2.6)$$
con $\varepsilon =2mV_0/\hbar$. All'aumentare di $d$ 
il temine in parentesi tender\`a a due, e quindi $\Delta\tau_\Trm^\varphi$ e 
$\Delta\tau_\Rrm^\varphi$ tenderanno a $2m/\hbar k\kappa=2/v_\grm\kappa$: tale valore
non dipende affatto dallo spessore della barriera
e diminuisce all'aumentare di $\kappa$, 
cio\`e per energie molto piccole ($k\rightarrow 0$).\par
Aumentando dunque lo spessore, possiamo trovare delle velocit\`a di 
tunnelling ${d/\tau_t^\varphi}$ arbitrariamente grandi: effetto 
Hartman$[8]$--Fletcher$[9]$ (o semplicemente effetto Hartman).
Tale effetto, per quanto potrebbe sembrare fisicamente 
inaccettabile, perch\'e in contrasto con la relativit\`a ed il principio di 
causalit\`a, \`e stato realmente osservato in tutti gli esperimenti cui abbiamo
precedentemente accennato.\par 
Quasi tutti gli autori$[10]$ insistono comunque sul
fatto che la violazione del principio di causalit\`a \`e, in questo caso,
solo apparente, e che il comparire di velocit\`a Superluminali nei risultati 
sperimentali sia in effeti dovuto ad un reshaping$^{\#3}$
\footnotetext{$^{\#3}$ Per reshaping del pacchetto intendiamo il fatto che, 
attraversando la barriera,
le sue componenti di energia pi\'u bassa sono, in genere, trasmesse meno 
efficientemente di quelle  di energia pi\'u alta. Ci\`o pu\`o provocare un 
cambiamento della forma del pacchetto nello spazio delle $k$ (una sua 
accelerazione in quello delle $x$), pi\'u o meno evidenti.}
del pacchetto.\\

\

Figura I-3.\\

Didascalia della Fig.I-3:\hfill\break
\{3a) Coefficienti di trasmissione $T(k)$ calcolati per diversi valori 
dello spessore $d$ di una barriera di potenziale rettangolare di altezza
$V_0=\hbar^2\varepsilon^2/2m$. Nella stessa figura viene inoltre riportato 
il grafico di $f(k-k_0)=e^{-(k-k_0)^2/2(\Delta k)^2}$, con 
$k_0=0.7\varepsilon$, $\Delta k=0.1k_0$.\hfill\break
3b) Coefficiente di trasmissione di una 
barriera di spessore $d=4/\varepsilon$ fino ad energie corrispondenti a 
valori di $k/\varepsilon =5$.\} \\

\

\h A questo punto, partendo proprio da un'analisi del \pht ,
pu\`o essere interessante una discussione, puramente qualitativa, 
degli effetti subiti da tutto l'insieme di frequenze costituenti un pacchetto
d'onda nell'attraversamento della barriera, per capire se, ed in quali casi, 
possano effettivamente apparire effetti dovuti al reshaping, 
e se sia possibile evitarli. \par
Tale discussione, che non pretende affatto di avere la validit\`a di 
una dimostrazione, trover\`a, comunque, conferma nei risultati ottenuti 
seguendo il centroide o l'evoluzione temporale del pacchetto, e si rende 
necessaria perch\'e, come $\tau_{\Trm,\Rrm}^\varphi$ e 
$\Delta\tau_{\Trm,\Rrm}^\varphi$, quasi tutte le altre definizioni di tempi
saranno calcolate in $k=k_0$.\ppar
In Fig.I-3a
possiamo vedere rappresentati i valori di $T(k)$, calcolati per diversi
valori di $d$ in funzione di $\varepsilon$.
Nello stesso grafico \`e riportata la funzione di 
distibuzione di un pacchetto gaussiano piccato intorno a $k_0=0.7\varepsilon$.
$f(k-k_0)=e^{-(k-k_0)^2/2(\Delta k)^2},$
con $\Delta k=0.1 k_0 .$ Naturalmente 
il peso di ogni componente di Fourier nel pacchetto trasmesso sar\`a dato
dal prodotto $T(k)f(k-k_0)$.\ppar
Per barriere molto sottili
$(d\varepsilon \ll 1)$ $\;T(k)$ \`e quasi costante
e molto vicino ad 1, tranne che per valori di $k$ molto piccoli. Purch\'e
dunque $k_0/\varepsilon$ non sia troppo vicino a $0$, la funzione 
$T(k)f(k-k_0)$ avr\`a il suo massimo sempre in $k_0$ ed il pacchetto uscente 
non presenter\`a deformazioni.\par
In questo caso
il tempo di trasmissione, calcolato sempre
come \pht , risulta maggiore del \teqz e non ci sono problemi. 
\ppar

Aumentando lo spessore della barriera, questa comincer\`a a trasmettere
sempre ``peggio" le componenti del pacchetto di energia pi\'u bassa: infatti,
per valori di $d$ abbastanza grandi,
$T(k)$ rimane molto piccolo tranne quando 
$k/\varepsilon $ si avvicina molto ad 1 dove, invece, cresce molto 
velocemente.\par 
Proprio per questo motivo, secondo molti autori, il pacchetto subirebbe 
un'accelerazione in quanto il massimo del pacchetto trasmesso
si verrebbe a trovare verso un valore $k_0^\prime > k_0$.
Dunque, secondo questi autori, \`e come se ad attraversare la barriera 
fossero solo un 
sottoinsieme di frequenze, quelle pi\'u elevate, propagatesi a velocit\`a
maggiore di $v_\grm$, anche prima di raggiungere la barriera stessa.
In effetti, \`e difficile che ci\`o avvenga, e pu\`o essere in ogni caso
evitato (senza che scompaiano le velocit\`a di gruppo Superluminali). Vediamo 
come.\par
Innanzitutto notiamo che la barriera non pu\`o avere alcun 
effetto amplificante su nessuna componente del pacchetto, ma solo fare da filtro
per alcune di esse, quindi $T(k)$ \`e limitata e
pu\`o salire solo fino ad un valore massimo pari ad
uno (come vediamo anche dalle Figg.I-3a e I-3b).\par
Per $k<\varepsilon$, inoltre, $T(k)$ \`e strettamente crescente e,
nel caso che ci 
interessa, cio\`e per barriere abbastanza spesse, $T(k)$ cresce  molto 
velocemente solo per valori di $k$ prossimi ad $\varepsilon$, 
dove ha un flesso obliquo in cui passa da concava a convessa. Saranno inoltre:
$$
\lim_{k\to \varepsilon} T(k)={
		1\over{\sqrt{
			1+\displaystyle{\varepsilon^2d^2\over 4}}}},
\eqno(I.2.7)$$
$$\lim_{k\to \varepsilon} T^\prime (k)=
   {{2d^2\varepsilon(3+d^2\varepsilon^2)}\over
					{3(1+d^2\varepsilon^2)}^{3\over 2}}
\eqno(I.2.8)$$ 
limiti validi anche per $k>\varepsilon$. \parn 
In prossimit\`a di $\varepsilon$
sar\`a quindi:
$$T(k)\simeq {
		1\over{\sqrt{
			1+\displaystyle{\varepsilon^2d^2\over 4}}}} +
   {{2d^2\varepsilon(3+d^2\varepsilon^2)}\over
					{3(1+d^2\varepsilon^2)}^{3\over 2}}
					(k-\varepsilon)
\eqno(I.2.9)$$
e dunque, nel punto in cui cresce pi\'u velocemente, $T(k)$ cresce non pi\'u 
velocemente di $T^\prime (k)(k-\varepsilon)$.\ppar
Per quanto riguarda $f(k-k_0)$ abbiamo invece che: 
$$f(k-k_0)=exp[-(k-k_0)^2/2(\Delta k)^2].$$\par
L'argomento dell'esponenziale sar\`a minore di uno per $(k-k_0)<\Delta k$ e
maggiore di uno se $(k-k_0)>\Delta k$: nel secondo dei due casi
$f(k-k_0)$ decrescer\`a
esponenzialmente con $((k-k_0)/\Delta k)^2$.\ppar
Tranne che per barriere troppo sottili, 
per le quali abbiamo gi\`a visto che $T(k)\simeq 1$,
ci aspettiamo che, se 
$(\varepsilon -k_0)\sim \Delta k$, buona parte del tunnelling possa anche 
essere dovuto alla trasmissione delle 
frequenze di energia pi\'u alta e, quindi, il picco delle $k$ potrebbe
effettivamente spostarsi in avanti: sar\`a per\`o, in ogni caso, 
$(k^\prime_0 - k_0) \sim \Delta k$. \par
Diminuendo, invece, l'energia del pacchetto incidente, 
prendendo cio\`e dei 
$k_0$ minori, il contributo al tunnelling da parte delle energie pi\'u alte 
diminuisce sensibilmente in quanto $f(k-k_0)$ decresce esponenzialmente con
$((k-k_0)/\Delta k)^2$: ci\`o sempre pi\'u velocemente man mano che $k_0$ 
si allontana da $\varepsilon$.\par 
Affinch\'e  sia possibile un reshaping del pacchetto, tale da provocare 
uno shifting o il formarsi di un secondo picco in avanti,
\`e allora necessario che, in certo intervallo, $T(k)$ cresca molto 
velocemente, 
abbastanza pi\'u velocemente di quanto decresce $f(k-k_0)$ in quello stesso
intervallo. \parn Per l'esattezza dovr\`a essere: 
	$${\drm\over \drm k}[T(k)f(k)]=
		T^\prime (k)f(k-k_0)+T(k)f^\prime (k-k_0)>0,
\eqno(I.2.10)$$ 
funzione molto difficile da studiare a causa della forma abbastanza complicata 
di $T^\prime(k)$.\parn
Poich\'e per\`o $f^\prime (k-k_0)=-\displaystyle{(k-k_0)\over (\Delta k)^2}f(k-k_0)$,
ed $f(k-k_0)>0$ sempre, la condizione (I.2.10) si riduce a:
$$T^\prime (k)-{(k-k_0)\over (\Delta k)^2}T(k)>0.
\eqno(I.2.11)$$
Osserviamo allora che,
come del resto ci aspettiamo, per $k<k_0$ la (I.2.11) \`e sempre verificata;
per $k>k_0$ invece, 
affinch\'e il prodotto $T(k)f(k-k_0)$ sia crescente, dovr\`a essere:
$$T^\prime (k)>T(k){(k-k_0)\over (\Delta k)^2}. \eqno(I.2.12)$$
Ma, come abbiamo gi\`a visto, $T^\prime (k)$ \`e limitata, e per barriere 
abbastanza spesse ha il suo massimo in $k=\varepsilon$.
Quindi,  per quanto possa essere piccola $T(k)$, possiamo sempre trovare dei 
valori di $\Delta k$ tali che la (I.2.12) non sia 
pi\'u valida e quindi il picco rimanga in $k_0$. 
Ci\`o anche per barriere molto spesse,$^{\#4}$
\footnotetext{$^{\#4}$ Per $\kappa d =(\varepsilon^2-k^2)^{1/2} d\gg 1$, si ha
$T(\kappa)=T((\varepsilon^2-k^2)^{1/2})\sim \displaystyle{4k\kappa\over
\varepsilon^2} e^{-\kappa d}$, ma in tal caso \`e anche $T^\prime (k)\sim 
e^{-\kappa d}$}
per le quali per\`o la probabilit\`a di tunnelling diviene infinitesima; basta 
solo che: \ 1) $k_0$ non sia troppo vicino a $\varepsilon$, \ e  \
2) la distribuzione delle $k$ sia abbastanza stretta intorno 
a $k_0$.\\

\

Figura I-4.\\

Didascalia della Fig.I-4:\hfill\break
\{Grafici di $T(k)f(k-k_0)$ per energie al di sotto dell'energia di
barriera in funzione del numero d'onda $k$, per diversi valori di $k_0$
e diversi valori dello spessore $a$ della barriera \
($k_0=0.3\varepsilon$, Fig.I-5a; \ $k_0=0.5\varepsilon$, Fig.I-5b; \
$k_0=0.7\varepsilon$, Fig.I-5c; \ $k_0=0.9\varepsilon$, Fig.I-5d; \
$\Delta k=0.1k_0$; \ e dall'alto verso il basso: \
$a=1/\varepsilon$, $a=3/\varepsilon$, $a=6/\varepsilon$, 
$a=10/\varepsilon$, $a=15/\varepsilon$, $a=20/\varepsilon$, 
$a=25/\varepsilon$. \ Notiamo che i grafici relativi agli ultimi due valori
non compaiono nelle prime due figure a causa della forte attenuazione del
picco.\} \\

\

\h In Fig.I-4 sono riportati i grafici di $T(k)f(k-k_0)$, per diversi valori dello 
spessore della barriera e di $k_0$ (notare la scala logaritmica).
Come si pu\`o vedere: solo per valori di $k_0/\varepsilon$ abbastanza vicini 
ad uno ($k_0/\varepsilon =0.9$) buona parte del tunnel avviene per componenti 
di energia al di sopra dell'energia di barriera.\ppar
Notiamo infine che, anche quando il picco dovesse spostarsi in avanti, ci\`o 
non 
avrebbe alcuna influenza diretta sulla sua posizione nello spazio delle $x$, se 
non fosse che, in tal caso, sia $\tau_{\Trm,\Rrm}^\varphi$, che 
$\Delta\tau_{\Trm,\Rrm}^\varphi$,
perderebbero qualsiasi significato fisico 
in quanto calcolati in $k=k_0$. 
\ppar
Tutto ci\`o ci assicura 
che {\em il pacchetto trasmesso, che andiamo a rivelare, non presenta delle 
grosse deformazioni rispetto a quello incidente}.\\

\

{\bf 
I.3) Tempi costruiti seguendo il centroide.}\\ \rm

Alcuni autori, anzich\'e seguire il picco attraverso il 
metodo della fase stazionaria, preferiscono usare come
riferimento il centroide (o centro di massa)
del pacchetto. Ci\`o, sia perch\'e, come abbiamo visto,
per poter applicare il metodo della fase stazionaria abbiamo bisogno di 
pacchetti molto stretti in $k$, e questo li rende molto estesi nello spazio 
ordinario, sia soprattutto perch\'e, cos\'{\i} facendo, \`e possibile valutare 
l'effetto di eventuali accelerazioni causate dall'attraversamento della 
barriera. \ppar
Supponiamo che inizialmente $(t\leq 0)$ il pachetto si trovi 
ad una certa distanza dalla barriera tale che:
$$\int_0^\infty |\psi (x,0)|^2 \drm x \simeq 0$$
che sia, cio\`e, trascurabile la probabilit\`a che 
la particella si possa venire a 
trovare oltre lo 0. Identifichiamo la posizione della particella all'istante 
$t=0$ attraverso la posizione del suo ``centro di massa", cio\`e:
$$\bar x(0)=\displaystyle{{\int_{-\infty}^\infty
			x|\psi(x,0)|^2 dx}\over
			{\int_{-\infty}^\infty |\psi(x,0)|^2 dx}}
\eqno(I.3.0)$$
Dato:
$$f(k)=f(k,0)\displaystyle{1\over\sqrt{2\pi}}\int dx\psi (x,0) e^{-ikx} =
      |f(k)|e^{i\xi(k)}
\eqno(I.3.1)$$
si pu\`o dimostrare che$[11]$:
$$x_0=\bar x(0)={{-\int_0^\infty \drm k|f(k)|^2\displaystyle{{\drm \xi}\over{\drm k}}}\over
{\int_0^\infty \drm k|f(k)|^2}}=-<\xi^\prime (k)>
\eqno(I.3.2)$$
e ponendo:
$$<...>=\int_{-\infty}^\infty \drm k|f(k)|^2;\qquad
<...>_{\Inrm}={{<...1>}\over{<1>}}=<...1>;$$
$$<...>_\Trm={{<...|T|^2>}\over{<|T|^2>}};\qquad
<...>_\Rrm={{<...|R|^2>}\over{<|R|^2>}}$$
avremo:
$$\bar x_{\Inrm}(t)=x_0+{\hbar\over m}<k>_{\Inrm} t,\qquad(t\to 0);\eqno(I.3.3)$$
$$\bar x_\Trm(t)=x_0+{\hbar\over m}<k>_\Trm t - <\alpha^\prime>_\Trm,
\qquad(t\to \infty);\eqno(I.3.4)$$
$$\bar x_\Rrm(t)=x_0+{\hbar\over m}<k>_\Rrm t - <\beta^\prime>_\Rrm,
\qquad(t\to \infty);\eqno(I.3.5)$$
Facendo evolvere in avanti nel tempo $\bar x_{\Inrm}(t)$ e $\bar x_\Rrm(0)$, ed 
indietro $\bar x_\Trm(t)$, si possono estrapolare $t_{\Inrm}(0)$, 
$t_\Rrm(0)$ e $t_\Trm(d)$, rispettivamente
come tempi per i quali il centroide passa per 0 e per $d$.$^{\#5}$
\footnotetext{$^{\#5}$ Notiamo che stiamo sempre usando delle forme 
asintotiche della funzione d'onda, trascurando quindi, sempre, gli effetti 
di autointerferenza in prossimit\`a della barriera.}
\ppar
I tempi di trasmissione e di riflessione saranno dati allora da:
$$\tau_\Trm^\Crm=t_\Trm(d)-t_{\Inrm}(0)={m\over\hbar}
	\Bigl[{1\over{<k>_\Trm}}\Bigl(d-x_0+<\alpha^\prime >\Bigr)+
				{{x_0}\over{<k>_{\Inrm}}}\Bigr];\eqno(I.3.6)$$
$$\tau_\Rrm^\Crm=t_\Rrm(0)-t_{\Inrm}(0)={m\over\hbar}
	\Bigl[{{-x_0+<\beta^\prime >_\Rrm}\over{<k>_\Rrm}}+
				{{x_0}\over{<k>_{\Inrm}}}\Bigr].\eqno(I.3.7)$$
\par 
Leavens e Aers$[12]$ dimostrano che, nel limite $\Delta k\to 0$,
si ha: $\tau_\Trm^\Crm \to \Delta\tau_\Trm^\varphi$ e
$\tau_\Rrm^\Crm \to \Delta\tau_\Rrm^\varphi$, e che, comunque, le eventuali correzione 
sono del primo ordine in $\Delta k$, il che conferma i ragionamenti di natura
qualitativa precedentemente fatti.\ppar
Risultati analoghi sono raggiunti anche da Martin e Landauer$[13]$ che 
eseguono il calcolo nel caso elettromagnetico, e da
Collins, Lowe e Barker$[3]$ che 
seguono l'evoluzione temporale di un pacchetto gaussiano attraverso l'equazione 
di Schr\"odinger time-dipendent. \par
Anche loro per\`o, pur calcolando esplicitamente il momento in cui il 
centroide lascia la barriera (non c'\`e autointerferenza per la funzione 
d'onda per $x\geq d$), {\em estrapolano} quello in cui questa viene raggiunta dal
centroide$[6]$.  \ppar
Passiamo ora ad esaminare alcuni dei tempi definiti mediante ``orologi".\\

\

{\bf 
I.4) Tempi di B\"uttiker e Landauer.}\\ \rm

Per determinare $\tau_\Trm$,
B\"uttiker e Landauer$[14-16]$, nel 1982, propongono di considerare una 
barriera rettangolare oscillante nel tempo, 
supponendo il tempo di attraversamento della sudetta barriera
pari al tempo di interazione delle particelle con il potenziale oscillante.\par 
Consideriamo allora
una barriera di potenziale rettangolare di altezza $V_0$, a cui 
venga sovrapposto un potenziale oscillante $\delta V \cos \omega t$.\par
A frequenze piuttosto basse, il potenziale varier\`a molto lentamente, dunque 
la particella, nell'attraversamento, risentir\`a solo di una parte del ciclo di 
modulazione e, finch\'e il periodo corrispondente all'oscillazione sar\`a 
lungo rispetto al tempo di attraversamento, la particella interagir\`a con un 
potenziale quasi statico.\par 
A frequenze pi\'u alte $(\omega\gg 1/\tau_\Trm)$, la particella, pur vedendo un 
potenziale medio $V_0$, subir\`a l'effetto di vari cicli di oscillazione e 
potr\`a assorbire o cedere dei quanti di energia pari a $\hbar\omega$.\par
La frequenza a cui avviene la transizione tra il comportamento adiabatico, 
tipico delle basse frequenze, e quello in cui si presenta assorbimento o 
cessione di energia, fornisce una misura del tempo di interazione della 
particella con la barriera. Naturalmente questa sar\`a soltanto un'indicazione
approssimata di tale tempo.\ppar
Al primo ordine in $\delta V$ appariranno solo le due bande $E\pm \hbar\omega$.
Inoltre le particelle appartenti alla banda energetica pi\'u alta saranno
favorite nell'attraversamento rispetto a quelle di energia pi\'u bassa.\par
B\"uttiker e Landauer mostrano 
quindi che, per barriere spesse e frequenze non troppo 
elevate ($\hbar\omega$ piccolo rispetto sia ad $E$ sia a $V_0-E$), 
l'intensit\`a relativa delle due bande sar\`a data da:
$$I_\pm^\Trm(\omega)={{|T_\pm (\omega )|}\over{|T(\omega )|^2}}
		 =\Bigl({{\delta V}\over{2\hbar\omega}}\Bigr)^2
		   \bigl[e^{\pm\omega {md\over\hbar \kappa}}-1\bigr]^2.
			\eqno(I.4.0)$$
I due autori identificano quindi il tempo di attraversamento con:
$\tau_\Trm^{\BLrm}=\displaystyle{md / (\hbar \kappa)}$.\parn
Nel limite di frequenze molto piccole la (I.4.0) si riduce a:
$$\Bigl|{{T_\pm (\omega )}\over{T(\omega )}}\Bigr|^2
		 =\Bigl({{\delta V}\over{2\hbar\omega}}\Bigr)^2.
					\eqno(I.4.1)$$\par
Come ci aspettiamo, il numero di particelle che avranno
assorbito o emesso energia \`e, in questo caso,
assolutamenete indipendente da $\omega$.\parn 
Sempre dalla (I.4.0) si ha :
$${T_+-T_-\over T_++T_-}=\tanh (\omega\tau_\Trm^{\BLrm})\eqno(I.4.2)$$\par
Ci\`o mostra come proprio $\tau_\Trm^{\BLrm}$ determini il passaggio dal 
comportamento adiabatico a basse frequenze, dove $T_+\simeq T_-$, a quello ad 
alte frequenze dove $T_+\gg T_-$.\ppar
Per quanto riguarda le particelle riflesse, sempre nei limiti
$\hbar\omega\ll E~~e\quad\hbar\omega\ll V_0-E$, trovano:
$$\Bigl|{R_\pm\over R}\Bigr|^2=\Bigl({\delta V \tau_\Rrm\over 2\hbar}\Bigr)^2,
\eqno(I.4.3)$$ con $\tau_\Rrm=\displaystyle{\hbar k /  (V_0\kappa)}$.\parn
Notiamo che anche la (I.4.3) \`e indipendente da $\omega$.\ppar
Il comportamento a basse frequenze di eq.(I.4.1) ed eq.(I.4.3)
\`e tipico di un sistema a due stati, $|\ 1>$ e $|\ 2>$, di energia $E$ ed 
$E\pm\hbar\omega$, portati in risonanza da una perturbazione 
$V_1\cos \omega t$.
Se per $t=0$ l'intera popolazione del sistema si trova nello stato $|\ 1>$, la 
popolazione dello stato $|\ 2>$ cresce inizialmente con 
$\displaystyle{\Bigl({V_1 t\over 2\hbar}\Bigr)}^2$. Analogamente avviene se 
le energie dei due livelli $E_1$ ed $E_2$ sono uguali. Nelle (I.4.1) e (I.4.3)
$\tau_\Trm^{\BLrm}$ e $\tau_\Rrm^{\BLrm}$ giocano quindi effettivamente il ruolo dei tempi
di interazione del sistema particella-barriera.\\

\

{\bf 
I.5) Tempi di Larmor.}\\ \rm

Nel 1966 Baz'$[17,18]$ propone di sfruttare la precessione di Larmor, 
causata dalla  
presenza di un campo magnetico su particelle dotate di spin,
per misurare i tempi di collisione di queste particelle.\par
Nello stesso anno Rybachenko$[19]$ 
applica questo metodo per calcolare i tempi di 
tunnelling nel caso unidimensionale di una barriera rettangolare.\ppar
Consideriamo allora un fascio di particelle di spin ${1\over 2}$, polarizzate 
in direzione $\hat x$, massa $m$ e
energia cinetica $E$, che si muovono in direzione $\hat y$ (vedi Fig.I-5). \
Supponiamo inoltre che  un campo magnetico omogeneo debole, 
$\vec B_0$, rivolto lungo 
l'asse $\hat z$, occupi la zona della barriera, sovrapponendosi a questa.\\

\

Figura I-5.\\

\

\h Seguendo il ragionamento di Rybachenco,
le particelle che penetreranno la barriera, attraversando il campo magnetico, 
subiranno una precessione di Larmor con frequenza $\omega_\Lrm=g \displaystyle
{\mu B_0 / \hbar}$, con $g$ rapporto giromagnetico, e $\mu$ momento 
magnetico delle particelle.\par
La precessione si arrester\`a nel momento in cui la particella riemerger\`a da 
una delle due facce della barriera.\parn
Poich\'e$[20]$ valgono $$<S_x>_\Trm={\hbar\over 2} \cos\omega_\Lrm\tau^\Lrm ,$$
$$<S_y>_\Trm=-{\hbar\over 2} \sin\omega_\Lrm\tau^\Lrm ,$$
nel limite di campi magnetici deboli avremo:
$$<S_x>_\Trm\simeq{\hbar\over 2} ,$$
$$<S_y>_\Trm\simeq-{\hbar\over 2}\omega_\Lrm\tau_{y\Trm}^\Lrm ,$$
dunque, la componente di spin acquistata in direzione $\hat y$ dalla particella
sar\`a proporzionale al suo tempo di permanenza all'interno della 
barriera. Poniamo allora:
$$\tau_{y\Trm}^\Lrm= \lim_{\omega_\Lrm\to 0} {<S_y>_\Trm\over {-{1\over 2}\hbar\omega_\Lrm}}$$
\par
Stranamente, per\`o, Rybachenko non prende in considerazione l'effetto maggiore 
del campo sulle particelle, cio\`e l'allineamento degli spin 
nella propria direzione. Uscendo dalla barriera, infatti,  
gli spin delle particelle 
avranno acquistato anche una componente lungo $\hat z$ pari a
$\pm\hbar/2$.\par 
Mentre al di fuori della barriera l'energia della particella era indipendente 
dallo spin, al suo interno questa dipender\`a anche dalla componente $z$ dello 
spin, a causa dell'effetto Zeeman. \par
La differenza  di energia per le particelle  spin-up e spin-down sar\`a 
$\pm\hbar\omega_\Lrm/2$. Le particelle spin-up penetreranno quindi meglio la 
barriera. Ponendo:
$$\psi_{\inrm}={1\over\sqrt 2}\left(\matrix{1\cr
					1}\right)e^{iky}$$ 
$$\psi_{\Trm}=(|D_+|^2+|D_-|^2)^{-1/2}\left(\matrix{D_+\cr
						 D_-}\right)e^{iky}$$
con:
$$D_\pm =T(\kappa_\pm)e^\alpha e^{-i\kappa_\pm d}$$
dove $\kappa_\pm$ \`e il $\kappa$ corrisponente a $E\pm\displaystyle{\hbar\over 
2}\omega_\Lrm$. 
Se:
$$ <S_\irm>_\Trm={\hbar\over 2} <\psi|\hat\sigma |\psi>, $$
otteniamo:
$$<S_z>_\Trm={\hbar\over 2}{|T_+|^2 -|T_-|^2\over |T_+|^2+|T_-|^2},\eqno(I.5.0a)$$
 
$$<S_y>_\Trm=-\hbar\sin (\alpha_+-\alpha_-){|T_+ T_-|\over |T_+|^2+|T_-|^2},
\eqno(I.5.0b)$$

$$<S_x>_\Trm=\hbar\cos (\alpha_+-\alpha_-){|T_+ T_-|\over |T_+|^2+|T_-|^2}.
\eqno(I.5.0c)$$\par
Espressioni analoghe si trovano per le particelle riflesse, basta sostituire 
$T_\pm$ con $R_\pm$. Si pu\`o inoltre dimostrare che:
$$<S_z>_\Rrm=-<S_z>_\Trm{|T_+|^2 -|T_-|^2\over |R_+|^2+|R_-|^2}, \eqno(I.5.1a)$$

$$<S_y>_\Rrm=-<S_y>_\Trm\left|{R_+R_-\over T_+T_-}\right|
{|T_+|^2 +|T_-|^2\over |R_+|^2+|R_-|^2}\eqno(I.5.1b)$$

$$<S_x>_\Rrm=-<S_x>_\Trm\left|{R_+R_-\over T_+T_-}\right|
{|T_+|^2 +|T_-|^2\over |R_+|^2+|R_-|^2}\eqno(I.5.1c)$$
Nel caso di campo magnetico debole 
$\kappa_\pm\simeq \kappa\mp \displaystyle{m\omega_\Lrm / \hbar},$
ed inoltre $$|T_+|^2 -|T_-|^2\sim -{m\omega_\Lrm\over\hbar\kappa}
			       {\partial T\over\partial\kappa}.
\eqno(I.5.2)$$\par
Nella (I.5.2) il termine $\displaystyle{m\over\hbar\kappa}
			     \displaystyle{\partial T\over\partial\kappa}$, 
che moltiplica $\omega_\Lrm$, ha, naturalmente, le dimensioni di un tempo.
B\"uttiker$[21]$ nel 1983
suggerisce allora 
l'introduzione di tre tempi $\tau_{z\Trm}^\Lrm$, $\tau_{y\Trm}^\Lrm$ e $\tau_{x\Trm}^\Lrm$,
nel seguente modo. Pone:
$$<S_z>_\Trm=(\hbar/2)\omega_\Lrm\tau_{z\Trm}^\Lrm, \eqno(I.5.3a)$$

$$<S_y>_\Trm=-(\hbar/2)\omega_\Lrm\tau_{y\Trm}^\Lrm, \eqno(I.5.3b)$$

$$<S_x>_\Trm=(\hbar/2)[1-(\omega_\Lrm^2{\tau_{x\Trm}^\Lrm}^2)/2], \eqno(I.5.3c)$$
saranno allora:
$$\tau_{z\Trm}^\Lrm=\lim_{\omega_\Lrm\to 0} {<S_z>_\Trm\over{{\hbar\over 2}\omega_\Lrm}}=
	 -{m\over\hbar\kappa}{\partial \ln T\over\partial\kappa},
\eqno(I.5.4a)$$

$$\tau_{y\Trm}^\Lrm=\lim_{\omega_\Lrm\to 0} {<S_z>_\Trm\over{{\hbar\over 2}\omega_\Lrm}}=
	 -{m\over\hbar\kappa}{\partial\alpha\over\partial\kappa},
\eqno(I.5.4b)$$

$$\tau_{x\Trm}^\Lrm=\lim_{\omega_\Lrm\to 0} {<S_z>_\Trm\over{{\hbar\over 2}\omega_\Lrm}}=
	 {m\over\hbar\kappa}\left[
		\left({\partial\alpha\over\partial\kappa}\right)^2 +
		\left({\partial \ln T\over\partial\kappa}\right)^2\right].
\eqno(I.5.4c)$$
Svolgendo i calcoli B\"uttiker ottiene:
$$\tau_{z\Trm}^\Lrm={m\varepsilon^2\over\hbar\kappa^2}
	{(\kappa^2-k^2)\sinh^2(\kappa d)+
		(\kappa d\varepsilon^2/2)\sinh (2\kappa d)\over
			4k^2\kappa^2+\varepsilon^4\sinh^2(\kappa d)}
\eqno5.5a$$
$$\tau_{y\Trm}^\Lrm={mk\over\hbar\kappa^2}
	  {2\kappa d(\kappa^2-k^2) + \varepsilon^2\sinh (2\kappa d)\over
		4k^2\kappa^2+\varepsilon^4\sinh^2(\kappa d)}
\eqno5.5b$$           
$$\tau_{x\Trm}^\Lrm=\sqrt{\tau_{z\Trm}^2+\tau_{y\Trm}^2}
\eqno5.5c$$
Per barriere abbastanza spesse si ha:
$$\tau_{z\Trm}^\Lrm\simeq {md\over\hbar\kappa},\qquad
		\tau_{y\Trm}^\Lrm\simeq {2mk\over \hbar\varepsilon^2\kappa}
\eqno(I.5.6)$$

Notiamo dunque che, nel limite di barriere spesse, $\tau_{z\Trm}^\Lrm=\tau_\Trm^{\BLrm}$.
Del resto, nel caso della componente $z$ dello spin, non abbiamo una vera e 
propria precessione, ma un ``salto" in posizione spin-up o spin-down, salto
accompagnato da uno splittamento dei livelli energetici. \par
Lo stesso B\"utticher$[21]$ dimostra inoltre che: 
anche per $\tau_{z\Trm}^\Lrm$ si pu\`o fare lo 
stesso discorso del sistema a due livelli fatto per $\tau_\Trm^{\BLrm}$. Questi due 
tempi comunque, a mio parere ed in base ai risultati sperimentali, 
vanno considerati come tempi di risposta della
particella alla perturbazione in qualche grado di libert\`a diverso dalla 
posizione, e quindi non come tempi di attraversmento.\par
Per quanto riguarda poi $\tau_{x\Trm}^\Lrm$, \`e ancora pi\'u difficile 
assegnargli un significato fisico: infatti, se immaginiamo che anche la 
componente $x$ dello spin preceda intorno all'asse $\hat z$, allora a tale 
precessione dovrebbe corrispondere una componente media dello spin in direzione
$x$ pari a:
$$<S_x>_\Trm=(\hbar/2)[1-(\omega_\Lrm^2{\tau_{y\Trm}^\Lrm}^2)/2],$$ e non a:
$$<S_x>_\Trm=(\hbar/2)[1-(\omega_\Lrm^2{\tau_{x\Trm}^\Lrm}^2)/2]. \eqno(I.5.3c)$$

Dunque alla luce di quanto detto prima su $\tau_{z\Trm}^\Lrm$,  $\tau_{x\Trm}^\Lrm$ 
potrebbe al massimo essere visto come una media di $\tau_{y\Trm}^\Lrm$ e
$\tau_{y\Trm}^\Lrm$ e, infatti, alcuni autori introducono direttamente un tempo
$\tau_\Trm^\Brm=\displaystyle{\sqrt{{\tau_{y\Trm}^\Lrm}^2 + {\tau_{y\Trm}^\Lrm}^2}}$. \parn
Nel limite di barriere spesse $\tau_\Trm^\Brm\simeq\tau_{z\Trm}^\Lrm$.\ppar
L'unico dei tre \tdlz qui definiti, cui sembra quindi possibile
attribuire un significato fisico, \`e $\tau_{y\Trm}^\Lrm$.\parn
Per questo, Falk e Hauge$[22]$ trovano, nel 1988, le seguenti due 
relazioni:
$$\tau_{y\Trm}^\Lrm={m\over\hbar k}(x_2-x_1+\alpha^\prime)+{mR\over 2\hbar k^2}
	[\sin (\beta -2kx_1)-\sin (2\alpha  - \beta + 2kx_2)]
\eqno(I.5.7a)$$
e 
$$\tau_{y\Trm}^\Lrm={m\over\hbar k}(x_2-x_1+\alpha^\prime)+{mR\over 2\hbar k^2}
	[\sin (\beta -2kx_1)-\sin (2\alpha  - \beta + 2kx_2)]+$$
		$${m\over 2\hbar k^2R}
			[\sin (\beta -2kx_1)+\sin (2\alpha  - \beta + 2kx_2)]
\eqno(I.5.7b)$$
per la parte riflessa , dove $x_1$ e $x_2$ sono due qualsiasi punti 
esterni alla barriera (uno alla sua destra ed uno alla sua sinistra)  
e il campo magnetico, anzich\'e essere limitato alla sola zona della barriera,
 \`e esteso a tutto l'intervallo ($x_1$,$x_2$).\ppar
Si vede facilmente che le (I.5.7) corrispondono ai \phtz pi\'u dei 
termini oscillanti, le cui ampiezze aumentano al diminuire dell'energia 
incidente.\parn
Nel caso di barriera rettangolare abbiamo:
$$\tau_{y\Trm}^\Lrm(d)=\tau_{y\Rrm}^\Lrm(d)=\tau^\Drm(d),$$
dove $\tau^\Drm$ non \`e altro che il \dwt , che introdurremo in seguito.\\

\

{\bf 
I.6) Tempi complessi: path-integral.}\\ \rm

L'introduzione di tempi complessi nasce inizialmente dalla seguente idea:
al di sopra dell'energia di barriera 
$v=\hbar\kappa=\hbar\displaystyle{\sqrt{k^2-\varepsilon^2}}$ e, dunque, 
$\tau_\Trm=\displaystyle{d\over v}=\displaystyle{md\over\hbar\kappa}$.\par
Per $E<V_0$, invece, il vettore d'onda diviene immaginario; consideriamo per\`o 
lo stesso il moto della particella sotto barriera, come se esso avvenisse 
lungo una traiettoria classica, ma a velocit\`a e tempo immaginari.\par
Naturalmente non \`e possibile dare a queste quantit\`a alcun significato 
fisico, in quanto, se ``dall'entrata" della particella all'interno della 
barriera, 
alla sua ``uscita" da una delle due faccie, sar\`a trascorso un certo tempo, 
per 
quanto piccolo (o grande) questo possa essere, dovr\`a per forza essere una 
quantit\`a reale.\par
Potremmo allora immaginare che siano le traiettorie ad essere complesse,   
ed  i tempi e le velocit\`a reali, considerando le quantit\`a:
$$v^\Srm={\hbar |\kappa |\over m},$$ 
		 $$\tau_\Trm^\Srm={d \over v}={md\over\hbar \kappa},$$ 
dove la $S$ sta per {\em semiclassiche}.\par
Naturalmente anche $\tau_\Trm^\Srm$ \`e fisicamente inaccettabile, in quanto diverge 
per $k=\varepsilon$ e resterebbe poi, in ogni caso, il problema delle 
particelle riflesse.\par
Notiamo comunque che,
nel caso di barriere spesse anche $\tau_t^{\BLrm}$, e naturalmente 
$\tau_{z\Trm}^\Lrm$, tendono a $\tau_\Trm^\Srm$.\par
Per quanto fisicamente inaccettabile, \`e interessante notare che nel 1992
Hagmann propone di considerare il caso di una particella che, per attraversare 
la barriera, acquisti una certa energia $\Delta E$, per un intervallo di tempo 
$\Delta t$. Il risultato che viene fuori applicando il principio di 
indeterminazione \`e proprio 
$\Delta t=\tau_\Trm^\Srm=\displaystyle{md\over\hbar \kappa}$.\ppar

Tornando ai tempi complessi, nel 1987, Sokolovski e Baskin$[23]$ propongono 
una generalizzazione del concetto classico di tempo alla meccanica 
quantistica ed applicano, poi, il metodo proprio al caso del tunnelling.\ppar
Cosideriamo una particella che, emessa all'istante $t_1$ nel punto $\vec r_1$,
sia rivelata all'istante $t_2$ nel punto $\vec r_2$. Supponiamo inoltre che la
particella, muovendosi lungo la traiettoria $\vec r(t)$, in un potenziale 
$V(\vec r)$, abbia attraversato una certa zona di spazio $\Omega$.\par
Il tempo trascorso dalla particella in quella regione di spazio sar\`a dato 
allora da:
$$\tau_{\clrm}^\Omega =\int_{t_1}^{t_2} \drm t \Theta_\Omega (\vec r(t)),\eqno(I.6.0)$$
dove $\Theta_\Omega (\vec r(t))$ \`e 1 se $\vec r(t)$ appartiene ad $\Omega$,
0 altrimenti. Nel caso unidimensionale avremo:
$$\tau_{\clrm}^\Omega =\int_{t_1}^{t_2} \drm t \int_0^d dx\  \delta (x-x(t)).
\eqno(I.6.1)$$\par
Se allora usiamo il metodo del {\em path-integral} di Feynman per costrurci 
delle traiettorie su cui mediare i tempi, otteniamo:
$$\tau^\Omega (x_1,t_1;x_2,t_2;k)=<\tau_{\clrm}^\Omega [x(.)]>_{\path},$$
in cui $x(.)$ \`e un cammino arbitrario (nello spazio delle fasi) tra
$(x_1,t_1)$ e $(x_2,t_2)$.
In generale $\tau^\Omega $ sar\`a complesso.\ppar
Sokolovski e Baskin trovano quindi:
$$\tau_\Trm^\Omega =i\hbar\int_0^d dx\ {\delta \ln A\over
					\delta V(x)},\eqno(I.6.2)$$

$$\tau_\Rrm^\Omega =i\hbar\int_0^d dx\ {\delta \ln B\over
					\delta V(x)},\eqno(I.6.2)$$
con $A=T e^{i\alpha }$,\ \ $B= R e^{i\beta }$. Nello stesso articolo i due autori
trovano una relazione tra $\tau^\Omega $ e i \tdl:
$$\Rerm \ \tau_\Trm^\Omega =\tau_{y\Trm}^\Lrm,\eqno(I.6.3a)$$
$$\Imrm \ \tau_\Trm^\Omega =\tau_{z\Trm}^\Lrm,\eqno(I.6.3b)$$
$$|\tau_\Trm^\Omega| =-\tau_{x\Trm}^\Lrm.\eqno(I.6.3c)$$
Analogamente per la parte riflessa avremo:
$$\Rerm \ \tau_\Rrm^\Omega =\tau_{y\Rrm}^\Lrm,\eqno(I.6.4a)$$
$$\Imrm \ \tau_\Rrm^\Omega =\tau_{z\Rrm}^\Lrm,\eqno(I.6.4b)$$
$$|\tau_\Rrm^\Omega| =\tau_{x\Rrm}^\Lrm.\eqno(I.6.4c)$$\par
Malgrado la connessione cos\'{\i} sorprendente tra $\tau^\Omega$ e i \tdl, \`e 
difficile  dare una interpretazione fisica di tali risultati, in quanto, come 
gi\`a detto, se esiste un tempo legato all'attraversamento della barriera, 
questo deve comunque essere reale.\par

\H Una (apparentemente un po' assurda, ma forse non tanto quanto sembra: si 
ricordi ad es. la teoria dell'elettrone, con cronone, di Caldirola: cfr.
R.H.A.Farias e E.Recami, ``Introduction of a quantum of Time (`chronon') and
its consequences for quantum mechanics", archivi elettronici LANL
\# quant-ph/9706059) \`e forse quella data da H\"anggi$[24]$ nel 1993. 
 \ Secondo
quest'autore il tempo di tunnelling sarebbe caratterizzato da due scale di 
tempi. Non sa per\`o spiegare dare un significato fisico all'idea.\par
Sokolovski e Connor,$[25]$ nello stesso anno, criticando tale affermazione,
insistono sul fatto che \`e impossibile e fisicamente insensato pensare 
che vi possano essere due tempi di 
attraversamento diversi. Concludono quindi che quello che va considerato come 
tempo 
di attraversamento \`e il modulo di $\tau_\Trm^\Omega$.\par
Soffermiamoci, per\`o, un attimo sull'ipotesi di H\"anggi.
Se guardiamo alla forma dell'onda trasmessa, possiamo osservare che:
$$\psi_\Trm=\psi_{\IIIrm}(x,k)=T(k)e^{i\alpha}e^{ikx}=e^{\ln T(k)}e^{i\alpha}e^{ikx}.
\eqno(I.6.5)$$\par
Notiamo allora che ogni qual volta applichiamo certi metodi
otteniamo un tempo
legato a $\displaystyle{\drm \alpha / \drm E}$, ed uno legato a 
$\displaystyle{\drm \ln T / \drm E}$.\par
Potremmo allora immaginare questo secondo tempo come 
il tempo che impiega la barriere a smorzare il segnale durante
l'attraversamento, tempo che non influirebbe affatto sulla sua velocit\`a.
Appare comunque strana, in questo caso, la dipendenza  di tale tempo da $d$, 
in quanto, in questo modo, $\tau_\Trm^\Omega$ anzich\'e 
diminuire all'aumentare dello spessore della barriera,
aumenta proporzionalmente a esso.\par
Andrebbe poi spiegato come tale tempo possa essere legato 
alle transizioni di livello, sia nel caso di B\"uttiker-Landauer, 
sia nel caso degli spin.\\

\

{\bf 
I.7) Tempi complessi: metodo di Bohm.}\\ \rm

Sempre nell'ambito dei tempi complessi, Leavens e Aers$[26]$, nel 1993, 
partendo dallo stesso operatore introdotto da Sokolovski e Baskin 
($\tau_{\clrm}^\Omega$), propongono l'uso del metodo di Bohm per ricavare delle 
traiettorie ``semiclassiche" da adoperare per il calcolo del tempo medio.\ppar
Il metodo di Bohm, come sappiamo, ci fornisce un'equazione del tutto 
equivalente
all'equazione di Schr\"odinger, consentendoci contemporaneamente 
un'interpretazione pi\'u ``classica"  dei fenomeni quantistici.
Accenniamo brevemente ad esso.\ppar
Poniamo $\psi= R e^{iS/\hbar}$, con $R$, $S$ reali. Applicando  
l'equazione di Schr\"odinger a $\psi$ e separando la parte reale e quella 
immaginaria, otteniamo una equazione di continuit\`a (la seconda delle due), 
ed un'equazione equivalente a quella di Hamilton-Jacobi per un potenziale 
$V(x)$ modificato, a cui \`e stato aggiunto il termine:
$$Q=-\left({\hbar^2\over 2m}\right) R^{-1}{\partial^2 R\over \partial x^2}$$
e la cui soluzione ci d\`a $S$.\par
La $R$ invece viene ricavata dall'equazione di continuit\`a.\par
Se consideriamo l'approssimazione WKB, e trascuriamo il potenziale quantistico
$Q$, otteniamo per la $S$ delle soluzioni di un'equazione di Hamilton-Jacobi 
classica nel potenziale originario: naturalmente, in questo caso, $S$ pu\`o 
essere, anzi sar\`a spesso, una quantit\`a complessa. Notiamo infine che, 
anche se quelle che otteniamo con il metodo di Bohm son traiettorie classiche, 
queste provengono da un potenziale modificato.\par 
I risultati fin qui ottenuti con questo metodo portano comunque a risultati 
fisicamente assurdi (vedi ref. [6] e [24]) e quindi non ce ne occupiamo.\\

\
       
{\bf 
I.8) Dwell time.}\\ \rm

Il \Dwtz viene introdotto per la prima volta da Smith$[27]$ nel 1960,
allo scopo di 
calcolare la durata media di un processo di collisione senza distinguere tra i 
vari canali e,
come gi\`a detto, viene definito come rapporto tra la 
probabilit\`a che la particella si trovi in una certa  regione di spazio 
ed il flusso $j_{\inrm}$ entrante in quella stessa regione, senza considerare, nel 
nostro caso, se la particella verr\`a riflessa o trasmessa:
$$\tau^\Drm(x_1,x_2;k)=j_{\inrm}^{-1}\int_{x_1}^{x_2}|\psi (x,k)|^2 dx=$$
	$$={1\over v_\grm}\int_{x_1}^{x_2}|\psi (x,k)|^2 dx. \eqno(I.0.0)$$
Nel caso di barriera rettangolare abbiamo:
$$\tau^\Drm(x_1,x_2;k)={mk\over\hbar\kappa^2}
	  {2\kappa d(\kappa^2-k^2) + \varepsilon^2\sinh (2\kappa d)\over
		4k^2\kappa^2+\varepsilon^4\sinh^2(\kappa d)}
\eqno(I.8.0)$$
che, per $\kappa d\gg 1$, diventa:
$$\tau^\Drm= {\hbar k\over V_0\kappa}={2mk\over h\varepsilon^2\kappa}
\eqno(I.8.1)$$\ppar

Per barriere abbastanza spesse, dunque,
anche $\tau^d$, come $\tau_{\Trm,\Rrm}^\varphi$ e $\Delta\tau_{\Trm,\Rrm}^\varphi$,
diventa indipendente dal loro spessore
ma, in forte contrasto con questi, diminuisce al diminuire di $k$, fino
ad annullarsi per $k=0$.\par
Data la sua definizione, ed il fatto che la trasmissione o la riflessione di 
una particella da parte di una barriera sono eventi mutualmente esclusivi, 
molti autori (quasi tutti) concludono che qualunque 
tempo candidato a rappresentare $\tau_\Trm$ e $\tau_\Rrm$ debba necessariamente 
verificare la relazione:
$$\tau_\Drm=|T(k)|^2\tau_\Trm + |R(k)|^2\tau_\Rrm. \eqno(I.0.1)$$ \par
Tale relazione \`e ad esempio verificata da $\tau^\Omega$. Sokolovski e 
Baskin$[23]$ nel 1987 trovano infatti la relazione:
$$\tau^\Drm=|R|^2\tau_\Rrm^\Omega +|T|^2\tau_\Trm^\Omega .$$
Da questa separando parte reale  e parte immaginaria si ottiene poi:
$$\cases{\tau^\Drm=|R|^2\tau_{y\Rrm}^\Lrm+|T|^2\tau_{y\Trm}^\Lrm\cr
	 |R|^2\tau_{z\Rrm}^\Lrm+|T|^2\tau_{z\Trm}^\Lrm=0.    \cr}$$\par
La prima di queste due equazioni viene ricavata indipendentemente anche da
Falck e Hauge$[22]$ nell'anno successivo.
La seconda, invece, non ci d\`a altro che una 
legge di conservazione del momento angolare, o meglio, in questo caso,
della componente $z$ dello spin.\par
La (I.01) invece sembra non essere verificata  nel caso dei \phtz e,
a maggior ragione, dei tempi di fase ``estrapolati". 
{\em Anzi}, Hauge et al.$[11]$ trovano la seguente relazione:
$$\tau^\Drm(x_1,x_2;k)=$$
$$|T(k)|^2\tau_\Trm^\varphi (x_1,x_2;k)+
		|R(k)|^2\tau_\Rrm^\varphi (x_1,x_2;k)+$$
			$${mR\over \hbar k^2}\sin (\beta -2kx_1)
\eqno(I.8.2)$$\par
Tuttavia, i due autori dimostrano anche che
se consideriamo che ogni pacchetto ha una certa larghezza in $k$ 
($\Delta k=\sigma$) e applichiamo la (I.8.2) a tutto il pacchetto, 
abbiamo:$^{\#6}$
\footnotetext{$^{\#6}$ Naturalmente in (I.8.3) dovremmo considerare 
$<|T(k)|^2\tau_\Trm^\varphi (x_1,x_2;k)>$ e $<|R(k)|^2\tau_\Rrm^\varphi
(x_1,x_2;k)>$, invece di $<|T(k)|^2><\tau_\Trm^\varphi (x_1,x_2;k)>$ e
$<|R(k)|^2><\tau_\Rrm^\varphi (x_1,x_2;k)>$. Si pu\`o per\`o dimostrare$[4]$ che 
l'errore che si commette usando la (I.8.4) \`e dell'ordine di $\sigma $.}
$$<\tau^\Drm(x_1,x_2;k)>\simeq$$
$$<|T(k)|^2><\tau_\Trm^\varphi (x_1,x_2;k)>+
		<|R(k)|^2><\tau_\Rrm^\varphi (x_1,x_2;k)>+$$
			$$+{mR\over \hbar k^2}
				\sigma^{-1}\int \drm k\sin (\beta -2kx_1)+
					O(\sigma )
\eqno(I.8.3)$$
Se dunque $|x_1|\gg \sigma^{-1}$, l'argomento dell'integrale osciller\`a 
abbastanza velocemente da renderlo trascurabile e quindi possiamo scrivere:
$$<\tau^\Drm(x_1,x_2;k)>=$$
$$<|T(k)|^2><\tau_\Trm^\varphi (x_1,x_2;k)>+
		<|R(k)|^2><\tau_\Rrm^\varphi (x_1,x_2;k)>
\eqno(I.8.4)$$
Relazione in accordo con la (I.0.1) a meno di $O(\sigma)$. \ppar
Naturalmente la (I.8.4) non \`e valida per i tempi di fase estrapolati, e 
mette nuovamente in evidenza il carattere puramente asintotico dei \pht.\ppar 
Prima di passare alle critiche mosse da  alcuni autori nei confronti del \dwt,
analizziamo un po' meglio la (I.8.2).\par
Secondo Hauge e Stovneng$[5]$ questa proverebbe che $\tau^\Drm$ rappresenta 
proprio il tempo esatto trascorso dalle particelle all'interno della barriera, 
mentre il termine $\displaystyle{mR\over \hbar k^2}\sin (\beta -2kx_1)$ 
dovrebbe 
rappresentare  un $\Delta \tau$ provocato dagli effetti di auto-interferenza.
\par
Infatti, se calcoliamo il \dwtz in un'intervallo $(-L,x_1)$ e facciamo tendere 
$L$ all'infinito, il valore che otteniamo diverger\`a anch'esso a $+\infty$.
Sottraendo per\`o il \dwtz calcolato in $(-L,x_1)$ solo sulla particella
incidente, otteniamo$^{\#7}$ un $\Delta\tau^\Drm(x<x_1;k)$
\footnotetext{$^{\#7}$ Un procedimento simile, in cui si valutano per\`o 
i flussi positivi e quelli negativi separatamente, verr\`a adottato da 
Olkhovsky e Recami per generalizzare la definizione di \dwt.} 
che, facendo i conti, risulta proprio: 
$$\Delta\tau^\Drm(x<x_1;k)=-{mR\over \hbar k}\sin(\beta -2x_1).\eqno(I.8.5)$$
La (I.8.2) pu\`o quindi essere riscritta come:
$$|T(k)|^2\tau_\Trm^\varphi (x_1,x_2;k)+
		|R(k)|^2\tau_\Rrm^\varphi (x_1,x_2;k)=$$
			$$\tau^\Drm(x_1,x_2;k)+\Delta\tau^\Drm(x<x_1;k)=$$
		$$\tau^\Drm(x_1,x_2;k)-    
			{mR\over \hbar k^2}\sin (\beta -2kx_1)
\eqno(I.8.6)$$\par

A favore di questa tesi i due autori portano il fatto che
il suddetto termine di 
auto-interferenza \`e del tutto indipendente  da $T(k)$ e da $\alpha (k)$, 
proprio perch\'e non c'\`e interferenza per $x>d$.
Aggiungono quindi gli esempi di due casi limite. \par
Il primo riguarda una barriera infinitamente spessa (in pratica uno scalino).
Essendo la barriera infinitamente spessa, non ci saranno particelle trasmesse e 
tutte le particelle verranno riflesse: $R=1$. \`E allora facile dimostrare che:
$$\Delta \tau_\Rrm^\varphi ={2\over\kappa v}={2m\over \hbar k\kappa}$$
$$\tau^\Drm={E\over V_0}\Delta\tau_\Rrm^\varphi ,$$
$$\Delta\tau^\Drm= {E-V_0\over V_0}\Delta\tau_\Rrm^\varphi .$$\par
In questo caso scompare l'apparente contraddizione tra il \tes, che aumenta con 
$k^{-1}$ al tendere di $k\to 0$, e il \dwtz che invece va a 0 con 
$E/v_\grm\sim k$. Infatti, se $\tau^\Drm$ \`e il tempo speso all'interno della 
barriera e $\Delta\tau^\Drm$ il ritardo (o anticipo) dovuto all'autointerferenza,
poich\'e al diminuire dell'energia incidente la particella penetrer\`a sempre
meno all'interno della barriera, a prevalere sar\`a il secondo termine, cio\`e 
$\Delta\tau^\Drm$.\par
Il secondo esempio riguarda, invece, una barriera a forma di delta di Dirac. 
Questo caso \`e uno dei primi ad essere trattati e risolti nella letteratura 
sull'argomento. Naturalmente, in questo caso, il \dwtz
sar\`a nullo, ma si ha:$^{\#8}$
\footnotetext{$^{\#8}$ In questo caso $d$ ha solo la funzione di un 
parametro, in quanto la $\delta$ di Dirac \`e costruita come limite di 
barriere sempre pi\'u strette ed alte ma di area costante $V_0d$.}
$$\Delta \tau_\Rrm^\varphi =\Delta\tau_\Trm^\varphi = T(k){V_0d\over mv^3},$$
con:
$$T(k)={1 \over {1+\left({V_0d\over\hbar k}\right)^2}}.$$
\noindent
Il tempo di tunnelling non pu\`o quindi che  provenire dal termine di 
autointerferenza.\\

{\bf 
I.9) Generalizzazione del dwell time.}\\ \rm

Come gi\`a accennato non tutti gli autori concordano sull'importanza fin
qui assegnata al \dwtz anche sulla base,  ma non solo, della (I.0.1).
Tale relazione, infatti, invocata come conseguenza del principio di 
sovrapposizione, e della mutua esclusivit\`a degli eventi di trasmissione o di 
riflessione, implicherebbe:
$$\int_{\rm Barriera}|\psi (x,k)|^2 dx\ =\ j(|T(K)|^2\tau_\Trm +|R(k)|^2\tau_\Rrm)=$$
$$= |T(K)|^2\ j\tau_\Trm +|R(k)|^2\ j \tau_\Rrm\eqno(I.9.0)$$
che quantomeno impone, a priori, $\tau_\Trm=\tau_\Rrm$, indipendentemente dalla
forma della barriera di potenziale.\ppar
Oltre a ci\`o, la (I.0.0) viene ricavata da B\"uttiker$[21]$
nel 1983 seguendo un 
metodo criticato, invece,  
da Olkhovsky e Recami.$[6,28,29]$ Infatti pur partendo da un'espressione
in cui appaiono esplicitamente le $\psi (x,t)$, nella definizione non si
terrebbe realmente in conto
della reale evoluzione temporale del pacchetto. Inoltre, a parte la relazione 
(I.0.1) la suddetta definizione non suggerisce nessun metodo per la separazione 
dei tempi dei diversi processi. 
Per questo motivo, nel tentivo di separare questi tempi, Olkhovsky e Recami
introducono preliminarmente le seguenti definizioni relative  ai tempi di 
trasmissione e di riflessione:
$$\overline{\tau_{\Trm}}=\overline{t(x_\frm)}_{\Trm}^{\IIIrm}-\overline{t(x_\irm)}_{\inrm}=$$
$${{\int_{-\infty}^{\infty} \drm t \, t\, J_{\Trm}^{\IIIrm}(x_\frm,t)}\over
	  {\int_{-\infty}^{\infty} \drm t \, J_{\Trm}^{\IIIrm}(x_\frm,t)}}
- {\int_{-\infty}^{\infty} \drm t\, t\, J_{\inrm}(x_\irm,t)\over
  \int_{-\infty}^{\infty} \drm t\,  J_{\inrm}(x_\irm,t)}=$$

$${\int_0^\infty \drm E \, v|g(E)T|^2 \tau_{\Trm}^{Ph} (x_{i},x_{f};E)\over
      \int_{0}^{\infty} \drm E\, v|g(E)T|^2}=$$

$$=(x_{f} -x_{i})< v^{-1}>_\Trm+<\delta \tau_\Trm >_\Trm\eqno(I.9.0a)$$
e
$$\overline{\tau_\Rrm}=\overline{t(x_\frm)}_\Rrm^\Irm-\overline{t(x_\irm)}_{\inrm}=$$

$$={{\int_{-\infty}^{\infty} \drm t \, t\, J_\Rrm^\Irm(x_\frm,t)}\over
	  {\int_{-\infty}^{\infty} \drm t \, J_\Rrm^\Irm(x_\frm,t)}}
- {\int_{-\infty}^{\infty} \drm t\, t\, J_{\inrm}(x_\irm,t)\over
  \int_{-\infty}^{\infty} \drm t\,  J_{\inrm}(x_\irm,t)}=$$

$${\int_0^\infty \drm E \, v|g(E)R|^2 \tau_{\Rrm}^{Ph} (x_{i},x_{f};E)\over
      \int_{0}^{\infty} \drm E\, v|g(E)R|^2}=$$

$$=(x_{f} -x_{i})< v^{-1}>_\Rrm+<\delta \tau_\Rrm >_\Rrm\eqno(I.9.0b)$$\par

Infatti, poich\'e
$J(x,t) \drm t$ rappresenta la densit\`a di probabilit\`a che una 
particella passi per il punto $x$, nell'intervallo di tempo $(t,t+ \drm t)$, per 
determinare il tempo medio in cui un pacchetto d'onda $\Psi (x,t)$ passa per il
punto $x$, dobbiamo fare la media della variabile $t$ pesata attraverso:
$$ w(x,t) ={J(x,t)\over\int_{-\infty}^\infty J(x,t) \drm t}.\eqno(I.9.1)$$ 

Subito dopo, per\`o, gli stessi autori$[6]$ notano che le definizioni (I.9.0) 
sono valide solo quando i 
pacchetti d'onda incidente e trasmesso sono completamente separati sia
spazialmente che temporalmente. Infatti quando $x_\irm$ ed $x_\frm$ non sono
abbastanza lontani dagli estremi della barriera  \`e possibile avere effetti di
interfernza tra la parte incidente e quella riflessa.\par
Inoltre, la densit\`a di corrente $J(x,t)$ pu\`o, in genere,
cambiare segno durante l'evoluzione temporale del pacchetto (per esempio
quando il picco dell'onda incidente raggiunge la barriera),
cosicch\'e  gli integrali
$\int_{-\infty}^{\infty} \drm t\, t J(x,t)$, che rappresentano la somma
algebrica di quantit\`a (flussi) positive e negative, e le densit\`a di
probabilit\`a $w(x,t)$, potranno non essere pi\'u definiti positivi.\par
In tal caso,
ciascuna densit\`a di probabilit\`a acquista un significato fisico
immediato solo durante gli intervalli di tempo in cui la corrispondente
densit\`a  di corrente  non cambia direzione:
occorre dunque spezzare l'integrale precedente in vari integrali, 
ciascuno dei quali sia considerato su un intervallo di tempo in cui il
segno di $J(x,t)$ sia solo positivo o solo negativo. In tal modo si
otterranno delle densit\`a di probabilit\`a tutte definite positive:
$$w_+(x,t)={J_{+}(x,t) \drm t\over \int_{-\infty}^{\infty} \drm t J_{+}(x,t) } $$
$$w_-(x,t)={J_{-}(x,t) \drm t\over \int_{-\infty}^{\infty} \drm t J_{-}(x,t) } $$
dove $J_{+}$ e $J_{-}$ rappresentano, rispettivamente, i valori
positivi e negativi di $J(x,t)$.\ppar

Alla luce di queste osservazioni, i due autori$[6]$ propongono come
tempi medi di trasmissione e di riflessione le seguenti espressioni:
$$\overline{\tau_\Trm}=\overline{t(x_\frm)}_+ -\overline{t(x_\irm)}_+ =$$
$${\int_{-\infty}^\infty  \drm t \, t J_+(x_\frm,t)\over
  \int_{-\infty}^\infty  \drm t\,  J_+(x_\frm,t)}
- {\int_{-\infty}^\infty  \drm t\, t J_+(x_\irm,t)\over
  \int_{-\infty}^\infty  \drm t\,  J_{+}(x_\irm,t)}
\eqno(I.9.2a)$$
e
$$\overline{\tau_\Rrm}=\overline{t(x_\irm)}_- - \overline{t(x_\irm)}_+ =$$
$$  {\int_{-\infty}^\infty  \drm t\, t J_{-}(x_\irm,t)\over
   \int_{-\infty}^\infty  \drm t\,  J_{-}(x_\irm,t)}
- {\int_{-\infty}^\infty  \drm t\, t J_{+}(x_\irm,t)\over
   \int_{-\infty}^\infty  \drm t\,  J_{+}(x_\irm,t)}
\eqno(I.9.2b)$$
Prima di passare oltre, partendo dall'equazione di continuit\`a:
$${\partial\rho (x,t)\over\partial t}+\, {\partial J(x,t)\over\partial x}\, =0,$$
e attraverso l'interpretazione probabilistica standard della $\rho (x,t)$, 
vogliamo dimostrare come le $w_\pm(x,t)$ appena definite corrispondano proprio 
alla probabilit\`a che la nostra particella, muovendosi in avanti o venendo
indietro, passi nell'intervallo di tempo $(t,t+ \drm t)$ per il punto $x$.\par
In ogni intervallo di tempo in cui $J=J_+$ o $J=J_-$, possiamo scrivere 
l'equazione di continuit\`a applicandola a $J_\pm$ [l'equazione di 
continuit\`a \`e in ogni caso valida sempre]:
$${\partial\rho_{{}_>}(x,t)\over\partial t}= 
		-{\partial J_+(x,t)\over \partial x}\eqno(I.9.3a)$$

$${\partial\rho_{{}_<}(x,t)\over\partial t}= 
		-{\partial J_-(x,t)\over \partial x} \ ,\eqno(I.9.3b)$$
ottenendo cos\'{\i} le due  quantit\`a, $\partial\rho_{{}_>}(x,t)/\partial t$ e
$\partial\rho_{{}_>}(x,t)/\partial t$. 
Integrando quindi rispetto al tempo in un intervallo $(-\infty, t)$ possiamo 
definire:
$$\rho_{{}_>}(x,t)=-\int_{-\infty}^t{\partial J_+(x,t)\over \partial 
x} \drm t^\prime\eqno(I.9.4a)$$

$$\rho_{{}_<}(x,t)=-\int_{-\infty}^t{\partial J_-(x,t)\over \partial 
x} \drm t^\prime\eqno(I.9.4a)$$
Imponiamo anche la condizione che $\rho_{{}_>}(x,-\infty )=0$, 
$\rho_{{}_<}(x,-\infty )=0$: supponiamo
cio\`e che inizilmente la particella (o il pacchetto 
d'onda) sia infinitamente lontano da $x$. Integrando nuovamente rispetto ad $x$
otteniamo altre due quantit\`a che indicheremo con $N_{{}_>}(x,\infty ;t)$, 
$N_{{}_<}(-\infty ,x;t)$ e per le quali si ha:
$$N_{{}_>}(x,\infty ;t)=\int_x^\infty \rho_{{}_>}(x^\prime ,t)dx^\prime =$$
$$\int_{-\infty}^t J_+(x,t^\prime ) \drm t^\prime >0,\eqno(I.9.5a)$$
$$N_{{}_<}(-\infty ,x;t)=\int_{-\infty}^x \rho_{{}_<}(x^\prime ,t)dx^\prime =$$
$$\int_{-\infty}^t J_-(x,t^\prime ) \drm t^\prime >0.\eqno(I.9.5b)$$\par 
Queste ultime due espressioni ci daranno la probabilit\`a che la nostra 
particella, muovendosi in avanti o all'indietro, si trovi, al tempo $t$, 
rispettivamente a destra o a sinistra del punto $x$ in 
funzione delle densit\`a di corrente $J_\pm (x,t)$.\par
Notiamo che le condizioni $\rho_{{}_>}(x,-\infty )=0$, 
$\rho_{{}_<}(x,-\infty )=0$ che avevamo imposto prima di integrare equivalgono 
ora a $J_\pm (-\infty ,t)=0$.\par
Finalmente, differenziando nuovamente le (I.9.5), questa volta  rispetto al 
tempo, abbiamo:
$$J_+(x,t)= {\partial\over\partial t} N_{{}_>}(x,\infty ;t)\, ,>0
\eqno(I.9.6a)$$
$$J_-(x,t)= {\partial\over\partial t} N_{{}_<}(-\infty ,x;t)\, ,>0
\eqno(I.9.6b)$$
e dunque:
$$w_+(x,t) ={\displaystyle{\partial\over \partial t}N_{{}_>}(x,\infty ;t)
	\over N_{{}_>}(x,-\infty ,\infty)}\eqno(I.9.7a)$$

$$w_-(x,t) ={\displaystyle{\partial\over \partial t}N_{{}_<}(-\infty ,x;t)
	\over N_{{}_<}(-\infty ,x,\infty)}.\eqno(I.9.7a)$$

Tali relazioni bastano a
giustificare la suddetta interpretazione probabilistica delle
$w_+(x,t)$ e $w_{-}(x,t)$.\ppar 
\ppar
A questo punto possiamo definire il
{\em valor medio} dell'istante  in cui la particella
passa per il punto $x$ muovendosi nella direzione positiva o negativa 
dell'asse. Saranno dunque:

$$\overline{t_+(x)}\ =\ {\int_{-\infty}^{\infty} t \, J_{+}(x,t) \drm t\over
		\int_{-\infty}^{\infty} J_{+}(x,t) \,  \drm t }\eqno(I.9.8a)$$

$$\overline{t_-(x)}\ =\ {\int_{-\infty}^{\infty} t \, J_{-}(x,t) \drm t\over
		\int_{-\infty}^{\infty} J_{-}(x,t) \,  \drm t }\eqno(I.9.8b)$$

Abbiamo ormai inoltre  tutti i mezzi per poter
definire anche le {\em varianze} delle distribuzioni 
relative ai suddetti tempi, e saranno:

$$\sigma^2(t_+(x)) ={\int_{-\infty}^{\infty} t^{2} J_{+}(x,t) \drm t \over
\int_{-\infty}^{\infty} J_{+}(x,t) \drm t} - (\overline{t_+(x)})^2
\eqno(I.9.9a)$$

$$\sigma^2(t_-(x)) ={\int_{-\infty}^{\infty} t^{2} J_{-}(x,t) \drm t \over
\int_{-\infty}^{\infty} J_{-}(x,t) \drm t} - (\overline{t_-(x)})^2
\eqno(I.9.9b)$$

Siamo quindi riusciti a costruire un formalismo che ci consente di
ricavare sia  i valori medi, che le varianze (o eventuali altri momenti),
relativi alle ``distribuzioni temporali" di tutti i possibili processi
relativi al tunnelling, nel caso unidimensionale.
Le stesse definizioni, comunque, sono estendibili a qualunque altro  
processo di collisione anche diverso dal tunnelling e con qualsiasi tipo di 
potenziale.\par
Come abbiamo gi\`a visto, per i tempi di tunnelling e di riflessione abbiamo:

$$\overline{\tau_\Trm}(x_\irm,x_\frm)=\overline{t(x_\frm)}_+ -\overline{t(x_\irm)}_+ =
\eqno(I.9.2a)$$
con $-\infty <  x_\irm  < 0$ e $d  <  x_\frm  < \infty $ e, in base alle
(I.9.9):
$$\sigma^2(\tau_\Trm(x_\irm,x_\frm))=\sigma^2(t_+(x_\frm)) +\sigma^2(t_+(x_\irm)).
\eqno(I.9.10)$$
Nel caso inolte $x_\irm=0$, $x_\frm=d$, abbiamo:
$$\cases{\overline{\tau_{{\rm Tun}}}(0,d)=\overline{t(d)}_+ -\overline{t(0)}_+\cr 
       \sigma^2(\tau_\Trm(0,d))=\sigma^2(t_+(d)) +\sigma^2(t_+(0)).        \cr}
\eqno(I.9.11)$$\par
Prendendo per esempio $x_\irm=0$ e $0<x_\frm<d$, possiamo ricavare tempi  di 
penetramento all'interno della barriera come:
$$\overline{\tau_{{\rm Pen}}}(0,x_\frm)=\overline{t(x_\frm)}_+ -\overline{t(0)}_+
\eqno(I.9.12)$$

o, per $0<x<d$:
$$\overline{\tau_{\rm Ret}}(x,x)=\overline{t(x)}_- -\overline{t(x)}_+
\eqno(I.9.13)$$

e, per $-\infty <x_\irm<d$:
$$\overline{\tau_\Rrm}(X_\irm,x_\irm)=\overline{t(x_\irm)}_+ -\overline{t(X_\irm)}_+.
\eqno(I.9.14)$$\ppar

Infine riesaminiamo, in base alla definizioni qui riportate, quelle
precedentemente date di \phtz e \dwt.\par
Per quanto riguarda il primo, salta per l'ennesima volta fuori il suo 
carattere puramente asintotico. Infatti, essendo 
questo ricavato in un contesto esplicitamente stazionario l'unica situazione in 
cui gli si pu\`o dare un senso fisico in base ai risultati qui riportati \`e 
quando $x_{\rm i} \to \infty$, quando cio\`e, $J_{+}(x,t)$ \`e la 
densit\`a di corrente del pacchetto iniziale in assoluta assenza di 
interferenza, tra la parte trasmessa e quella riflessa, 
dovuta alla onde riflesse.
Analogamente, il dwell time,  rappresentato dall'espressione 
equivalente:$[30,31]$
$$\overline \tau^{\Drm}(x_{\rm i},x_{\rm f}) =$$
$$ \left[
\int_{-\infty}^{\infty} t \; J(x_{\rm f},t) \; \drm t -
\int_{-\infty}^{\infty} t \; J(x_{\rm i},t) \; \drm t \; \right] \; \left[
\int_{-\infty}^{\infty} J_{\inrm}(x_{\rm i},t) \; \drm t \,
\right]^{-1}  \ ,$$
con $-\infty < x_{\rm i} < 0$, and $x_{\rm f}>d$, \ non \`e in genere
fisicamente significativo:
infatti il peso  nelle medie temporali \`e definito positivo, e normalizzato 
ad 1, solamente nei rari casi in cui
$x_{\rm i} \rightarrow-\infty$ and $J_{\inrm} = J_{\IIIrm}$ \ (i.e.,
quando la barriera \`e trasparente).\\

\

{\bf 
I.10) Tempi di penetrazione e di ritorno: risultati numerici.}\\ \rm

Le (I.9.2) e le (I.9.11-14) non consentono un calcolo analitico agevole 
per ricavare delle espresioni per i tempi di tunnelling (penetrazione, ritorno e 
riflessione), nemmeno nel caso alquanto semplice di barriera rettangolare.\par
Presentiamo dunque i risultati di alcuni calcoli numerici 
relativi alla durata media di vari processi di  penetrazione e ritorno 
di pacchetti gaussiani {\em all'interno} di una barriera rettangolare,
svolti da Olkhovsky et al. in un recente articolo$[29]$.
Tali calcoli, comunque,
confermano l'apparire dell'{\em effetto Hartman} e, in base a verifiche 
indirette, risultano, secondo gli autori, abbastanza
in accordo con i dati sperimentali di Colonia, Berkeley, Firenze, Vienna.\\     

\

Figura I-6.\\

Didascalia della Fig.I-6:\hfill\break
\{6a) Andamento del tempo di penetrazione 
medio, \tpen (espresso in secondi), 
in funzione della lunghezza di penetrazione $x_{\rm f} = x$
(espressa in  in {\AA}ngstrom) per una barriera rettangolare
di ampiezza $d = 5 \; {\rm \AA}$, con \
$\Delta k = 0.02 \; {\rm {\AA}}^{-1}$ (linea tratteggiata)
 \ e \ $\Delta k = 0.01 \; {\rm {\AA}}^{-1}$ (linea continua). \ Vale la pena
di notare che \tpen cresce rapidamente per i primi
 {\AA}ngstrom ($\sim 2.5$ \AA) iniziali, e tende poi
ad un valore di saturazione.
Questo sembra confermare l'esistenza del
cosiddetto ``effetto Hartman".\hfill\break
6b)Come in 6a) con $\Delta k = 0.01 \;
{\rm {\AA}}^{-1}$, \  ma con una barriera di spessore doppio
$d = 10 \; {\rm \AA}$. \ Osserviamo che
i valori numerici del tempo di tunnelling totale
\ttun restano praticamente inalterati quando si passa da
 \ $d = 5 \;$\AA  \ a \ $d = 10 \;$\AA, \ ad ulteriore conferma dell'apparire
dell'effetto Hartman.\} \\

\

\h Ricodiamo che, in base alle notazioni precedentemente riportate, \`e
$$\Psi_{\inrm}(x,t) = \int_{0}^{\infty}Cf(k-\overline k) \;
\exp[ikx-iEt/\hbar] \; \drm k $$
dove:
$$ f(k- \bar k) = e^{- {(k-\bar k)^2\over
2 (\Delta k)^2}} ,$$
$E = \hbar^{2} k^{2}/2m$,\ 
$C$ \`e una costante di normalizzazione ed $m$ \`e in questo caso
la massa dell'elettrone.
Le  lunghezze di penetrazione saranno espresse
in  {\AA}ngstroms, e il tempo di penetrazione in secondi.\\

\h In Fig.I-6a sono riportati i  grafici di
$\overline{\tau_{{\rm Pen}}}(0,x),\,\, 0<x<d$, 
corrispondenti a $d = 5 \;$\AA, \ per  \ $\Delta k = 0.02$ e
$0.01 \; {\rm {\AA}}^{-1}$. \ Come si pu\`o vedere il tempo di penetrazione
$\overline{\tau_{{\rm Pen}}}(0,x)$ tende sempre ad un valore di massimo di
{\em saturazione}.\par
In Fig.6b mostriamo invece il grafico corrispondente  a $d = 10\ $\AA,
$\Delta k = 0.01 \; {\rm {\AA}}^{-1}$. \`E interessante
notare che, a parit\`a  di $\Delta k$,  $\overline{\tau_{{\rm Pen}}}$ \`e
praticamente  lo stesso, sia  per  $d = 5$ che per
$d = 10 \;$\AA,  un risultato questo
che conferma ancora una volta l'apparire del cosiddetto
{\em  effetto Hartman.}.
Risultati analoghi sono stati ottenuti anche per $d >\ 10 \ {\rm \AA}$,
facendo  variare il parametro  $\Delta k$ tra  $0.005$
e le energie  $\overline E$ nel range da $1$ a $10$ eV.\\

\

Figura I-7.\\

Didascalia della Fig.I-7:\hfill\break
\{Andamento di \tpen (espresso in secondi) in funzione di
$x$ (in {\AA}ngstroms), relativo al  tunnelling attraverso una barriera 
di spessore $d = 5 \;${\AA} e per vari valori di ${\overline E}$ e di $\Delta k$: \ \
curva 1: \ $\Delta k = 0.02 {\rm \AA}^{-1}$ e $\overline E = 2.5 \; {\rm eV}$; \ \
curva 2: \ $\Delta k = 0.02 {\rm \AA}^{-1}$ e $\overline E = 5.0 \; {\rm eV}$; \ \
curva 3: \ $\Delta k = 0.02 {\rm \AA}^{-1}$ e $\overline E = 7.5 \; {\rm eV}$; \ \
curva 4: \ $\Delta k = 0.04 {\rm \AA}^{-1}$ e $\overline E = 5.0 \; {\rm eV}$.\} \\

\

\h Nelle  Figg.I-6, I-7, I-8 
sono riportati gli andamenti delle durate medie per i processi
di penetrazione e ritorno in funzione della lunghezza di penetrazione
(con  $x_{\rm i} = 0$ e $0\le \, x_{f} = x \le d$), per barriere
di altezza  $V_{0} = 10$ eV
e ampiezza $d= 5 \;${\AA} o in alcuni casi $10\; $\AA. \
In particolare:\hfill\break
--- In Fig.I-6, sono riportati i grafici di
$\overline{\tau_{{\rm Pen}}}(0,x)$ correspondenti a diversi
valori dell'energia cinetica media: \ \
$\overline E$ = 2.5 eV, 5 eV e 7.5 eV,  $\Delta k=0.02 {\rm \AA}$ 
(curve 1, 2 e 3); \ \
$\overline E = 5$ eV, $\Delta k =0.04 {\rm \AA}^{-1}$ (curva 4);\par
per tutti e quattro i casi \`e $d= 5 {\rm \AA}$.\hfill\break
--- In Fig.I-7 mostriamo i grafici di  $\overline{\tau_{{\rm Pen}}}(0,x)$
corrispondenti a: \ \
$d=5 {\rm \AA}$, con  $\Delta k = 0.02 {\rm \AA}^{-1}$ e 0.04 ${\rm \AA}^{-1}$
(curve 1 and 2); \ \ 
$d=10 {\rm \AA}$, con $\Delta k=0.02\ {\rm AA}^{-1}$
e 0.04 ${\rm \AA}^{-1}$ (curve 3 e 4,); \ \
dove l'enegia cinetica media $\overline E$ \`e 5 eV, e cio\`e met\`a 
dell'energia di barriera $V_{0}$.\hfill\break
--- Infine in Fig.I-8 mostriamo alcuni grafici di
$\overline \tau_{{\rm Ret}}(x,x)$. Le curve  1, 2 e 3 corrispondono
rispettivamente a: \ \
$\overline E = 2.5$ eV,  5 eV e 7.5 eV, per $\Delta k=0.02 {\rm \AA}^{-1}$ 
e $d = 5  {\rm \AA}$;\hfill\break
le curve  4, 5 and 6 corrispondono invece a: \ \ 
$\overline E = 2.5$ eV, 5 eV  e 7.5 eV, per
$\Delta k = 0.04 {\rm \AA}^{-1}$ e  $d = 5 {\rm \AA}$;\hfill\break
mentre le curve 7, 8 e 9 corrispondono rispettivamente a: \ \
$\Delta k =0.02 {\rm \AA}^{-1}$ e 0.04 ${\rm \AA}^{-1}$,
$\overline E = 5$ eV,  $d=10 {\rm \AA}$.\\

\

Figura I-8.\\

Didascalia della Fig.I-8:\hfill\break
\{Andamento di  \tpen (in secondi) in funzione di $x$ (in {\AA}ngstroms)  
per  $\overline E = 5 \; {\rm eV}$, e per vari valori  di $d$ e $\Delta k$: \ \
curva 1: \ $d = 5  \; {\rm \AA}$, $\Delta k = 0.02 {\rm \AA}^{-1}$; \ \
curva 2: \ $d = 5  \; {\rm \AA}$, $\Delta k = 0.04 {\rm \AA}^{-1}$; \ \
curva 3: \ $d =10 \; {\rm \AA}$, $\Delta k = 0.02 {\rm \AA}^{-1}$; \ \
curva 4: \ $d =10 \; {\rm \AA}$, $\Delta k = 0.04 {\rm \AA}^{-1}$.\} \\

\

\h Riguardo al modo in cui sono stati effettuati questi calcoli, gli autori fanno 
osservare che l'integrazione su  d$t$ \`e stata eseguita usando
l'intervallo temporale $[-10^{-13}, \ +10^{-13}]$ s, simmetrico
rispetto a $t = 0$. Tale intervallo risulta di tre ordini di grandezza
maggiore dell'estensione temporale del pacchetto d'onde che
ricordiamo essere dell'ordine di  $1/(\bar v \,
{\Delta k}) = ({\Delta k} \, \sqrt{2 \overline E / m})^{-1} \sim 10^{-16}$ s.
\par
Sottolineamo che ci\`o equivale a considerare l'evoluzione del pacchetto 
d'onda attraverso un intervallo temporale [$-\infty, \;\; +\infty$], e
non invece come se  questo iniziasse a evolversi ad un certo istante $t$ 
finito. Tutto ci\`o in accordo con le relazioni $J_\pm (-\infty ,t)=0$ o 
equivalentemente $\rho_{{}_<}(x,-\infty )=0$.
Aggiungiamo inoltre che il pacchetto \`e costruito in modo tale da arrivare 
con il centroide in $x_0$ al tempo $t=0$.\\  

\

Figura I-9.\\

Didascalia della Fig.I-9:\hfill\break
\{Andamento di \tret (in secondi) in funzione di  $x$ (in {\AA}ngstrom)  
per diversi valori di $d$,  $\overline E$ e $\Delta k$: \ \
curva 1: \ $d = 5 \; {\rm \AA}$, \ $\overline E = 2.5 \; {\rm eV}$ e
$\Delta k = 0.02 {\rm \AA}^{-1}$; \ \
curva 2: \ $d = 5 \; {\rm \AA}$, \ $\overline E = 5.0 \; {\rm eV}$ e
$\Delta k = 0.02 {\rm \AA}^{-1}$; \ \
curva 3: \ $d = 5 \; {\rm \AA}$, \ $\overline E = 7.5 \; {\rm eV}$ e
$\Delta k = 0.02 {\rm \AA}^{-1}$; \ \
curva 4: \ $d = 5 \; {\rm \AA}$, \ $\overline E = 2.5 \; {\rm eV}$ e
$\Delta k = 0.02 {\rm \AA}^{-1}$; \ \
curva 5: \ $d = 5 \; {\rm \AA}$, \ $\overline E = 5.0 \; {\rm eV}$ e
$\Delta k = 0.02 {\rm \AA}^{-1}$; \ \
curva 6: \ $d = 5 \; {\rm \AA}$, \ $\overline E = 7.5 \; {\rm eV}$ e
$\Delta k = 0.02 {\rm \AA}^{-1}$; \ \
curva 7: \ $d = 5 \; {\rm \AA}$, \ $\overline E = 5.0 \; {\rm eV}$ e
$\Delta k = 0.02 {\rm \AA}^{-1}$; \ \
curva 8: \ $d = 5 \; {\rm \AA}$, \ $\overline E = 5.0 \; {\rm eV}$ e
$\Delta k = 0.02 {\rm \AA}^{-1}$.\} \\

\

\h Dalle suddette Figure I-7)--I-9)  si pu\`o vedere che:

1) la durata media del processo di tunnelling
$\overline{\tau_{{\rm Tun}}}(0,d)$  non dipende dalla profondit\`a $d$ della
barriera (``effetto Hartman");  

2) la quantit\`a
$\overline{\tau_{{\rm Tun}}}(0,d)$ decresce all'aumentare dell'energia, come nel caso 
dei \tes;  

3) Il valore della durata del tempo di penetrazione
aumenta rapidamente soprattutto nella parte iniziale della 
barriera, in prossimit\`a  cio\`e del punto $x= 0$;

4) $\overline{\tau_{{\rm Pen}}}(0,x)$ tende a un valore di saturazione della parte 
finale della barriera, cio\`e per $x\to d$.\ppar
Per quanto riguarda gli effetti riportati nei punti  3) e 4), 
secondo gli autori, sarebbero causati dall'interferenza
tra le onde  iniziali che penetrano nella barriera
e le onde che tornano indietro (sempre dentro la barriera) e la cui 
sovrapposizione produce i flussi  $J_{+}$ e $J_{-}$. \ Si vedano anche le 
illustrazioni (in particolare Fig.3, pag.351) in ref.$[6]$. \par
Infatti, essendo nella parte iniziale della barriera
il pacchetto d'onda di ritorno abbastanza grande, ci\`o che accade  \`e che
il suo flusso estingue sostanzialmente buona perte di quello entrante del
pacchetto incidente. \par
All'aumentare di  $x$, per\`o, il pacchetto d'onde di ritorno va a zero 
pi\'u velocemente di quello entrante: ci\`o provocherebbe un aumento
della durata media del processo di penetrazione $\overline{\tau_{{\rm Pen}}}(0,x)$, 
facendo crescere velocemente tale durata, soprattutto nella zona iniziale della 
barriera.
Nella regione finale della barriera il suo aumento svanisce rapidamente
e si ha l'effeto inverso.\par

Infine, in connessione con i grafici
di  $\overline{\tau_{{\rm Ret}}}(x,x)$ in funzione di $x$, mostrati in  Fig.I-9,
osserviamo che: 

5) la durata media di riflessione
$\overline{\tau_{\Rrm}}(0,0)=\overline{\tau_{{\rm Ret}}}(0,0)$ \
non dipende dall'ampiezza  $d$ della barriera;

6) in corrispondenza della regione della barriera tra  $0$ e, circa, $0.6\, d$
il valore di  $\overline{\tau_{{\rm Ret}}}(0,x)$ \`e quasi costante;

7) il suo valore aumenta con $x$ solamente nella regione della barriera
vicino ad  $x = d$, anche se come sottolineano gli autori i calcoli relativi 
a tale regione non sono molto accurati a causa del fatto che la quantit\`a \
$\int_{-\infty}^{\infty} J_{-}(x,t) \drm t$ \
assume valori molto piccoli).

\h Notiamo quindi che il punto 5), previsto, come il punto 1)
per particelle quasi-monocromatiche 
da Dumont e Marchioro,$[32]$ risulta in accordo con i dati ottenuti da
Steinberg et al.$[33]$ per pacchetti d'onde arbitrari.
Come prima anche i punti 6) e 7) possono essere spiegati da fenomeni di
interferenza all'interno della barriera: infatti, se vicino ad  $x=d$
il pacchetto d'onde di ritorno iniziale \`e smorzato quasi totalmente
dal pacchetto d'onda incidente iniziale, allora rester\`a solamente una parte
trascurabile della sua coda posteriore (fatta dalle componenti con velocit\`a
minori).\par
Al diminuire di  $x$\ ($x \to 0$) la parte non smorzata del pacchetto di 
ritorno sembra diventare sempre pi\'u grande
(riacquistando le componenti pi\'u rapide)
cosicch\'e la differenza
$\overline{\tau_{{\rm Ret}}}(0,x) - \overline{\tau_{{\rm Pen}}}(0,x)$ rimane, con buona 
approssimazione, costante. 
Inoltre l'interferenza tra le onde incidenti e riflesse
nei  punti $x \le 0$ costituisce in effetti un fenomeno ritardante
cosicch\'e $\overline{t_-(x=0)}$ \`e pi\'u grande
di  $\overline{\tau_{\Rrm}}(x=0)$, \ il che pu\`o spiegare
i valori pi\'u grandi di $\overline{\tau_{\, \rm R}}(x=0,x=0)$  rispetto a 
$\overline{\tau_{{\rm Tun}}}(x=0,x=a)$.

\vfill \newpage

\centerline{{\bf Parte II: ESPERIMENTI.}}

\

\

{\bf 
II.0) Equivalenza ottica del tunnelling.}\\ \rm
Come gi\`a accennato, varie conferme sperimentali  
dell'{\em effetto Hartman} si sono recentemente avute in seguito ad una serie 
di misure effettuate a Colonia$[34]$, Berkeley$[35]$, 
Firenze$[36]$, e Vienna$[37]$.\par
Tali misure, per\`o, sono state eseguite sui tempi di trasmissione di microonde
e fotoni, sfruttando la propagazione di modi evanescenti all'interno di 
guide d'onda sotto la frequenza di cut-off, nel primo caso, e la riflessione 
frustrata, nel secondo.\par
Notiamo infatti che, malgrado in passato siano state avanzate numerose proposte
di esperimenti da effettuare direttamente con particelle quali ad esempio 
elettroni, grosse difficolt\`a erano state incontrate dal punto di vista 
pratico nel realizzarli, a causa soprattutto  dei tempi molto piccoli coinvolti 
in tali processi di tunnelling.\par
Per una giunzione Josephson, ad esempio, tali tempi 
risultano essere di qualche decina di {\em femto-secondi}, e possono  scendere
all'ordine del fs in altri dispositivi a stato solido.\par
Nel caso di sistemi ottici, invece, tali tempi sono gi\`a dell'ordine di 
qualche ps, per frequenze nella regione del visibile, e raggiungono
il ns
in alcuni degli esperimenti compiuti a Firenze ed a Colonia, con microonde.
\ppar
Pur senza prendere qui in rassegna tali esperimenti, oramai noti e 
famosi$[38]$, 
ed i risultati da questi ottenuti, occupiamoci dell'equivalenza tra 
trasmissione di modi elettromagnetici evanescenti e tunnelling 
di particelle, in particolare nel caso delle guide d'onda.
Consideriamo una particella di massa $m$ ed energia cinetica 
$E=\hbar^2k^2/ 2m$. Nel caso 
(unidimensionale) di attraversamento di un potenziale uniforme $V_0$, 
l'equazione {\em di Schr\"odinger} per tale particella sar\`a:
$${\partial^2\psi\over\partial x}+{2m\over\hbar^2}(E-V_0)\psi =0.\eqno(II.0.0)$$
Posto allora:

$$\kappa^2={2m\over\hbar^2}(E-V_0),\eqno(II.0.1)$$

la (II.0.0) risulta formalmente identica all'equazione {\em di Helmholtz}
per la componente scalare relativa al campo elettrico, o a quello magnetico,
di un campo (e.m.)  che si propaghi in un mezzo dispersivo:
$${\partial^2\psi\over\partial x}+\kappa^2\psi =0,\eqno(II.0.2)$$
dove in questo caso:
$$\kappa ={2\pi\over\lambda _m}={2\pi\over\lambda }n,$$
$\lambda_m\;$ \`e la lunghezza d'onda all'interno del mezzo, $\lambda\;$ \`e la
lunghezza d'onda nel vuoto,  e $n$ \`e l'indice di rifrazione del 
mezzo  in cui il campo si propaga.\par
Il confronto tra le due equazioni suggerisce  la 
sostituzione: 
$$\sqrt{{2m\over\hbar^2}(E-V_0)}\rightarrow {2\pi\over\lambda }n.$$
Nel caso di una guida d'onda rettangolare di dimensioni $a\times b$ ($a<b$), e 
con pareti perfettamente conduttrici, sappiamo che:\\ 
$$\kappa ={2\pi\over\lambda }\sqrt{1-\left({\lambda \over 2b}\right)^2}=
	{2\pi\over\lambda }\sqrt{1-\left({\lambda \over\lambda _c}\right)^2}
\eqno(II.0.3)$$
dove $\lambda _c=2b$ \`e la lunghezza d'onda di \em cut-off \rm al di sopra 
della quale il termine sotto radice divene negativo e, di conseguenza, $\kappa$ 
immaginario.\par
Poich\'e inoltre $\lambda  =c/\nu =2\pi c/\omega$, abbiamo:
$$\kappa =\sqrt{{\omega^2\over c^2} -{\pi^2\over b^2}}=
	{\omega\over c}\sqrt{1-\left({\omega_c\over\omega}\right)^2}
\eqno(II.0.4)$$
con $\omega_c=\pi c/b$ ($\nu_c = c/2b$),
che ci d\`a a sua volta la frequenza di \em cut-off
\rm al di sotto della quale $\kappa$ diviene immaginario.\par
Notiamo che la relazione di dispersione nel caso di guida d'onda rettangolare
\`e sorprendentemente uguale$[39]$ (si vedano anche le ``Feynman 
Lectures"$[7]$) a quella di una particella relativistica quando 
facciamo la sostituzione:
$${\pi\over b}={\omega_c\over c}= {mc\over \hbar}$$
infatti, moltiplicando la (II.0.4) per $\hbar$ abbiamo:
$$\hbar\kappa =\sqrt{{(\hbar\omega)^2\over c^2} -{(2\hbar\pi )^2\over b^2}}
	\rightarrow p^2=\sqrt{{E^2\over c^2}-mc^2}$$\par
Onde evitare confusione, precisiamo che qui il confronto lo stiamo facendo tra 
l'equazione di Helmholtz, che \`e relativistica e classica (perch\'e ricavata 
dalle equazioni di Maxwell), e l'equazione di Schr\"odinger, quantistica e 
{\em non} relativistica.\par
Proprio questo \`e, inoltre, uno dei vantagi di usare campi elettromagnetici 
anzich\'e particelle.
Torniamo, dunque, al suddetto confronto tra la (II.0.0) e la (II.0.2).
Differenziando la (II.0.1) e la (II.0.4), otteniamo:
$$v_{{\rm gruppo}}^{{\rm particella}}={\drm\omega\over\drm\kappa}={\hbar\over m}\kappa$$
$$v_{{\rm gruppo}}^{{\rm c.em}}={\drm\omega\over\drm\kappa}={c^2\over \omega}\kappa.$$
Attraverso la sostituzione:
$${\hbar\over m}\; \rightarrow\; {c^2\over\omega}={c\over 2\pi\nu},
\eqno(II.0.6)$$
possiamo sempre adattare i risultati elettromagnetici, nel caso di trasmissione 
di microonde in una guida d'onda, a quello quantistico, del moto 
unidimensionale di una particella in un potenziale uniforme.$^{\#9}$
\footnotetext{$^{\#9}$ Notiamo che la (II.0.6) equivale alla relazione 
$\hbar\omega =mc^2$}.
In entrambi i casi infatti le soluzioni delle due equazioni saranno date da 
combinazioni lineari delle funzioni d'onda:
$$\psi (x,t)=e^{\pm i\kappa x}e^{i\omega t}$$\par
Inoltre, quando l'energia della particella risulta minore di 
$V_0$, o quando nel caso elettromagnetico la pulsazione $\omega$ diviene minore 
di $\omega_c$, $\kappa$ diviene immaginario e le funzioni d'onda divengono 
degli esponenziali decrescenti$^{\#10}$ della forma $e^{-|\kappa |x}$
\footnotetext{$^{\#10}$ Le 
soluzioni del tipo $e^{|\kappa |x}$, per motivi fisici, si riferiscono
al set-up sperimentale opposto.} 
(onde evanescenti).\par
Notiamo dunque che il campo (o equivalentemente la funzione d'onda della 
particella) penetra ugualmente all'interno della zona sotto \coff (o 
classicamente proibita, per le particelle) almeno per una distanza dell'ordine
di $|\kappa|^{-1}$.\par 
Ovviamente resta il fatto che, nonostante le analogie formali,
vi sono delle differenze  fisiche tra il tunnelling di elettroni
e la propagazione guidata di microonde sotto \cof. Infatti, come gi\`a 
detto, l'equazione di Helmholtz (per le onde) e l'equazione di
Schr\"odinger (per elettroni) sono si, sostanzialmente, la stessa equazione, ma
mentre nel primo caso ci\`o  che si propaga \`e effettivamente una componente 
del campo (e quindi \`e il campo stesso), nel secondo caso si 
tratta della funzione d'onda della particella.\par 
Tuttavia, trattandosi in ambo i casi dell'evoluzione di pacchetti d'onda,
nulla ci vieta di interpretare i risultati degli esperimenti su microonde
come delle vere e proprie ``simulazioni fisiche" del tunnelling di elettroni, 
anche 
perch\'e il fatto che i risultati di tali simulazioni$[34-38]$ 
riproducano abbastanza bene le previsioni quantistiche conferma l'equivalenza 
tra i due casi.\par
Pi\'u sottile \`e
la circostanza che nel caso dipendente dal tempo le equazioni di Schroedinger
(in presenza di barriera) e l'equazione di Helmholtz (per onde 
elettromagnetiche in guida d'onda) non sono pi\'u matematicamente
equivalenti, dato che la derivata temporale \'e del primo ordine in un caso, 
e del secondo ordine nell'altro caso. \ Ciononostante, si pu\`o far 
vedere$[40]$ che esse ammettono ancora classi di soluzioni analoghe, 
differenti solo per le loro propriet\`a di spreading.\\

\

{\bf 
II.1) Misure del gruppo di Colonia.}\\ \rm

Riportiamo ora i risultati di alcune delle misure eseguite a Colonia dal gruppo di 
Nimtz, Enders, et al.[34],  effettuate  sulla trasmissione di microonde all'interno 
di una guida d'onda, al di sotto della {\em frequenza di cut-off}.\\

\

Figura II-1.\\

Didascalia della Fig.II-1.:\hfill\break
\{Configurazioni sperimentali del gruppo di Colonia.\} \\

\

\h Pi\'u precisamente, sono state utilizzate due guide d'onda rettangolari (ved.
Fig.II.1), la prima delle quali, di dimensioni 10.16\xx 22.86 $mm^2$ (banda 
$X$), \`e stata sempre utilizzata  a frequenze sempre superiori a quelle  di 
cut-off, di 6.56 GHz.\par
A met\`a di questa era inserita una seconda guida d'onda, di dimensioni 
7.90\xx15.80 $\rm mm^2$ (banda $Ku$), la cui frequenza di cut-off era di 9.49 
GHz.\par
Onde evitare eventuali errori sistematici provenienti dal restringimento dovuto 
alla connessione delle due guide, o all'interferenza tra la linea di 
``iniezione" e di ricezione del segnale con la guida 
d'onda di banda $X$, gli autori hanno utilizzato una speciale tecnica di calibrazione 
(calibrazione LRM). \  [Pi\'u recentemente \`e 
apparso un nuovo lavoro degli stessi autori in cui l'apparato sperimentale \`e 
stato leggermente variato; infatti, per evitare eventuali problemi dovuti alla 
strozzatura nel passaggio tra le due guide di banda $X$ e $Ku$,
anzich\'e usare una seconda guida per 
creare la zona di evanescenza, sono state inserite all'interno di questa degli 
strati di materiale  con indice di rifrazione diverso, sfuttando cos\'{\i} la 
riflessione frustrata, come nel caso degli 
esperimenti con fotoni: i risultati di tali nuove misure concordano 
perfettamente con quelli qui riportati.] \\

\

Figura II-2.\\

Didascalia della Fig.II-2:\hfill\break
\{Ritardi di fase sperimentali del 
coefficiente di trasmissione totale $T(\nu )$,
in funzione della frequenza, per quattro diverse lunghezze della guida d'onda 
centrale di banda $Ku$.\} \\

\h Una prima serie di misure \`e stata fatta nel dominio di frequenza, 
misurando cio\`e il coefficiente di trasmissione totale $T(\nu )$, al variare 
della frequenza portante di un certo inpulso (che considereremo pi\'u
avanti).\par  La quantit\`a $T(\nu )$ include quindi, in questo caso, anche
la variazione di fase causata  dall'attraversamento 
della  guida di banda $Ku$.\par
In effetti, le misure della fase e dell'ampiezza sono state eseguite 
separatamente, anche se contemporaneamente.

\h Le misure sono state effettuate nell'intervallo di frequenze: 8.2---9.2 GHz, 
appena 300 MHz al di sotto della frequenza di cut-off. Tale intervallo \`e 
stato diviso in 801 punti di misura: la differenza di frequenza tra due 
punti di misura successivi essendo di 1.25 MHz.

\h In Fig.II.2 sono riportati i ritardi di fase ottenuti per quattro 
differenti lunghezze della guida d'onda centrale; pi\'u esattamente \ $L \ = \
40,\ 60,\ 80,\ 100$ mm. 

\h \`E pi\'u che evidente che i suddetti ritardi di fase sono uguali per 
tutte e quattro le lunghezze con un'accuratezza di $\pm 1$ grado. Solo alle 
frequenze pi\'u piccole appaiono delle deviazioni statistiche maggiori.\\

\

Figura II-3.\\

Didascalia della Figura II-3:\hfill\break
\{In figura sono riportati l'impulso di riferimento 
(L=0 --- linea punteggiata); l'impulso trasmesso attraverso una 
``barriera"  di 100 mm (linea continua), con 
un ritardo di circa 130 ps; l'impulso trasmesso lungo la stessa distanza 
di 100 mm, ma attraverso una guda d'onda di banda $X$ \ (linea 
tratteggiata).\} \\

\

\h Va per\`o notato che a basse frequenze l'attenuazione del segnale \`e 
tale da far scendere la precisione degli strumenti di misura. 
Inoltre, anche nel vuoto un'onda elettromagnetica del genere di quelle usate 
presenta normalmente una variazione di fase dell'ordine di 1 grado, dopo aver 
viaggiato per una distanza di appena 0.1 mm. L'indipendenza dello sfasamento 
dalla lunghezza delle guide intermedie dimostra  che questo \`e causato 
essenzialmente solo dalle condizioni al contorno sulle due facce (d'ingresso e 
di uscita) della {\em guida d'onda} di banda $Ku$.\\

\

Figura II-4.\\

Didascalia della Fig.II-4:\hfill\break
\{Impulso di riferimento (L=0), linea continua; impulsi di tutte e 
quattro le  misure, che sono circa uguali, linee punteggiate;  impulso che 
attraversa sempre i 100 $mm$ ma in una guida di banda $x$ al di sopra quindi 
della frequenza di cut-off, linea tratteggiata.\} \\

\

\h Le stesse misure sono state riportate, dagli autori, nel dominio 
temporale attraverso un'integrazione di Fourier. Supponiamo infatti di avere 
un segnale $F(t)$, dato da :
$$F(t)=\int_{-\infty }^\infty f(\nu ) e^{i 2\pi\nu t} \drm t$$
dove $f(\nu )$ \`e la trasformata inversa di Fourier del segnale $F(t)$.
Dopo che il segnale avr\`a viaggiato attraverso una certa regione di spazio,
la sua nuova forma, nel dominio del tempo, sar\`a:
$$F^\prime (t)=\int_{-\infty }^\infty f(\nu )T(\nu ) e^{i 2\pi\nu t} \drm t .
\eqno(II.1.0)$$
\par
Dunque, una volta misurata $T(\nu )$, e conosciuta la forma del segnale 
iniziale, \`e possibile ricavare direttamente le misure dei 
tempi di arrivo del segnale. Gli autori adottarono come forma del 
segnale quella di una distribuzione di Gauss, supponendo, inoltre, che fosse 
possibile limitare l'integrale (II.1.0) all'intervallo di frequenze 
$(\nu_1,\nu_2)$, dove $\nu_1$ e $\nu_2$ sono le frequenze di \coff delle due 
guide d'onda. 

\h In Figg.II.3 e II.4 sono riportati, rinormalizzati, i risultati 
ottenuti trasformando le precedenti misure nel dominio del tempo (ovvero, 
in pratica, le $F^\prime (t)$). Notiamo come la deformazione degli impulsi 
attraverso la regione della barriera sia trascurabile: non c'\`e quindi 
assolutamente bisogno di distinguere tra picco e centro di massa 
dell'impulso.\\

\

Figura II-5.\\

Didascalia della Fig.II-5:\hfill\break
\{Inviluppi delle potenze dei segnali trasmessi per due frequenze 
portanti: a) {\rm 8.44\ GHz}, b) {\rm 8.658\ GHz}. Le figure continue 
rappresentano gli inviluppi dei segnali prima della barriera, attenuati di 
{\rm 40 dB}.\} \\

\

\h Gli stessi autori eseguono poi una seconda serie di misure direttamente 
sui tempi di attraversamento mediante un analizzatore di tempi. In questo 
caso,  il segnale viene modulato in ampiezza creando un 
treno di impulsi, con frequenza portante $\nu_0$. La durata di ogni impulso \`e 
di un $\mu$s, mentre la frequenza di ripetizione di 10 kHz.\par
L'analisi di Fourier del segnale comprende,
in questo caso, uno spettro infinito di frequenze, distribuite, comunque, 
intorno alla frequenza portante $\nu_0$. Tale frequenza va dunque scelta in 
modo da evitare che l'impulso trasmesso sia prevalentemente composto da 
frequenze al di sopra di quella di \coff. \par
Siccome per\`o queste frequenze non 
vengono attenuate, generano distorsioni abbastanza vistose del 
segnale.
Secondo gli autori, \`e abbastanza semplice, verificare quando ci\`o accade, e 
mettersi in condizioni da  poterlo evitare (cf. la prima delle refs.[34]).
\par
Le misure sono state fatte in questo caso sui tempi di arrivo del fronte d'onda 
di ogni singolo impulso.

\h In Figg.II.5a e II.5b, sono rappresentati (sia in scala naturale, a sinistra, 
sia in scala logaritmica, a destra), in funzione di $t$, gli inviluppi 
delle intensit\`a di due impulsi, con frequenze portanti rispettivamente di 
8.644 e 8.658 GHz, per uno spessore della guida centrale di 60 mm. Le linee 
continue invece rappresentano lo stesso inviluppo per il segnale di riferimento 
($L=0$). Quest'ultimo veniva fatto passare attraverso un attenuatore di 
potenza a scalino, che ne riduceva l'ampizza di 40 dB. L'errore temporale 
introdotto dall'attenuatore \`e comunque inferiore a 25 ps, indipendentemente 
dalla frequenza.  \par
Come si pu\`o vedere, anche in questo caso, il tempo di attraversamento risulta 
indipendente dallo spessore della guida ``subcritica" centrale.\par 
Misure analoghe sono state fatte, sempre dagli stessi autori, anche per altri 
valori della lunghezza della guida centrale, ottenendo risultati simili 
a quelli delle due figure qui riportate.\\

\

Figura II-6.\\

Didascalia della Fig.II-6:\hfill\break
\{Rappresentazione schematica della configurazione 
sperimentale nel caso della doppia barriera.\} \\

\

\h Il caso pi\'u interessante studiato dallo stesso gruppo, infine, 
\`e stato quello dell'attraversamento di una doppia barriera (cio\`e 
di due tratti di guida d'onda subcritica). Riportiamo qui
i risultati sperimentali ottenuti in questo caso, in quanto i tempi 
di attraversamento misurati, per bande di frequenze lontane dalle frequenze 
di risonanza, non solo risultano essere indipendenti dallo 
spessore delle due barriere, ma sono indipendenti anche dalla distanza tra 
queste due.\\

\

Figura II-7.\\
Didascalia della Fig.II-7:\hfill\break
\{La figura a destra riporta le
intensit\`a trasmesse di un pacchetto d'onda gaussiano 
in funzione del tempo nel caso di doppia barriera, per le tre configurazioni 
[a), b), c)] adottate. Notare l'equivalenza delle configurazioni a) e b).
 \  Sulla sinistra invece \`e riportato il coefficiente di 
trasmissione delle configurazioni a) e b). Per la configurazione b) si vedono
apparire due frequenze di risonanza.\} \\

\

\h In Fig.II.6, possiamo vedere le configurazioni sperimentali adottate:

\h a) dapprima, le due guide di banda $Ku$, entrambe di lunghezza 40 mm, 
sono affiancate l'una all'altra;

\h b) successivamente, una terza guida di banda $X$ e lunghezza 57
mm viene posta tra le due guide sottodimensionate;

\h c) la guida di 57 mm viene, infine, rimossa e montata in coda alla altre, 
mentre le due guide di banda $Ku$ vengono riaccostate.\\

Con la configurazione b) appaiono due frequenze di risonanza, rispettivamente 
a 6.9 e 7.6 GHz. In prossimit\`a di tali risonanze, i tempi di tunnelling
tunnelling misurati dagli autori tendono al valore inverso della larghezza 
della risonanza. \par  Pi\'u importante \`e la misura del tempo di 
attraversamento di un pacchetto gaussiano, stretto in $k$ \ ($\Delta k=0.01 k$), 
alla frequenza {\em non risonante}  di 7.3 GHz. I risultati di tale misura 
sono riportati in Fig.II.7 (grafico a sinistra). Come si pu\`o vedere, il 
tempo di 
attraversamento  del segnale \`e uguale sia nel caso a), senza ``buca di 
potenziale", sia  nel caso b) con una guida di banda $X$ di 57 mm posta tra 
le due guide di banda $Ku$. \par
Nel caso c) il tempo di attraversamento dei 57 mm risulta di 400 ps in 
accordo con la velocit\`a di gruppo all'interno della guida. \ppar 
Dalla seconda delle due Figure II.7, notiamo che il coefficiente di 
trasmissione, lontano dalle frequenze di risonanza, assume all'incirca gli 
stessi valori, sia nel caso a), sia nel caso b).

\h I due autori passano quindi a considerare ---teoricamente--- il caso
della misura con la configurazione b), ma con una guida intermedia  lunga
1 km (invece di 57 mm). In tal caso, naturalmente, il numero delle 
risonanze aumenter\`a, diminuendo la distanza tra l'una e l'altra. Il pacchetto 
d'onda usato per misurare il tunnelling non risonante va dunque stretto ad una 
larghezza di $10^{-6}$ volte la frequenza centrale, onde evitare effetti di 
interferenza. Allora,
per una frequenza portante di 10 GHz, sarebbe possibile, ad esempio, 
trasmettere  un segnale di larghezza 10 KHz. Secondo gli autori un segnale del 
genere basterebbe  per modulare e trasmettere anche la Sinfonia n.40 di Mozart
ad una distanza di 1 km ed a velocit\`a Superluminale. Gli autori si
riferiscono a simulazioni attraverso computer[41], di cui danno per\`o 
pochi dettagli.  Tali computer simulations sono state pertanto rifatte
con cura presso il D.M.O. della Electric Engineering Faculty della 
Campinas State University, Campinas, S.P., Brasile[42], con analoghi 
risultati di cui si daranno i risultati altrove.\\

\

Figura II-8.\\

Didascalia della Fig.II-8:\hfill\break
\{Grafici dei tempi di trasmissione calcolati da Hartman, in funzione dello 
spessore della barriera ($d\varepsilon$), e del numero d'onda $\kappa'$ 
rinormalizzato a $\varepsilon$ \ ($\kappa' \equiv \kappa/\varepsilon$, 
essendo $\varepsilon$ il numero 
d'onda corrispondente all'altezza della barriera). I punti segnati
rappresentano approssimativamente i dati sperimentali per un pacchetto 
elettromagnetico gaussiano, con frequenza centrale 8.7 GHz,  nel caso 
di spessori della guida centrale (frequenza di \coff 9.49 GHz) di 10, 
40, 60, 80, 100 mm.\} \\

\

\h Riportiamo, infine, i grafici dei tempi di tunnelling di una barriera di 
potenziale di spessore $d$ ed altezza $\hbar^2\varepsilon^2/2m$, calcolati da 
Hartman. Nello stesso grafico sono riportati i relativi 
{\em tempi equivalenti}. I punti segnati
rappresentano approssimativamente i dati sperimentali per un pacchetto
elettromagnetico gaussiano, con frequenza centrale 8.7 GHz,  nel caso
di spessori della guida centrale (frequenza di \coff 9.49 GHz) di 10,
40, 60, 80, 100 mm. \ Notiamo l'impressionante precisione dei dati 
sperimentali.\\

\

{\bf 
II.2) Misure del gruppo di Berkeley.}\\ \rm

Prendiamo ora in considerazione i famosi esperimenti condotti a
Berkeley dal gruppo di Chiao, Steinberg e Kwiat.\par
In questo caso le misure sono state svolte sfruttando la 
{\em band-gap} dovuta alle riflessioni frustrate su un materiale 
multistrato. Anche in questo caso, comunque, vale l'analogia tra 
l'equazione di Schr\"odinger e quella di Helmholtz, e possiamo estendere 
i risultati delle misure al caso di barriere di potenziale quantistiche. 
Quella che cambia sar\`a, al solito, la 
relazione di dispersione del mezzo e, come nel caso delle guide d'onda, 
anche qui, in un certo intervallo di frequenze, il vettore d'onda 
$\kappa$ diviene immaginario, e appaiono i modi evanescenti. \par
L'uso di fotoni, per\`o, presenta alcuni vantaggi,
rispetto a quello delle microonde: per esempio, il fotone dimostra un 
comportamento ``individuale" diverso da quello 
presentato da un segnale elettromagnetico qualsiasi nel campo delle microonde. 
Inoltre anche la dispersione \`e molto pi\'u ridotta, quasi come nel caso
di una particella.\\ 

\

Figura II-9.\\

Didascalia della Fig.II-9:\hfill\break
\{Configurazione sperimentale di Berkeley.\} \\

\

\h In Figura II.9 \`e schematizzato l'apparato sperimentale usato in questo 
caso.\par
Una coppia di elettroni viene generata attraverso un processo
di riconversione spontanea provocato da un laser 
all'interno di un cristallo (KDP) di potassio-di-idrogeno-fosfato.
I due elettroni sono generati contemporaneamente e mentre uno dei due viaggia 
liberamente in aria, l'altro viene fatto incidere su un 
supporto di materiale rifrangente, met\`a del quale  \`e ricoperto da un 
multistrato composto di due sostanze con indice di 
rifrazione diverso. Ogni strato ha uno spessore pari a un quarto della lunghezza 
d'onda dei fotoni. Lo spessore totale \`e di 1.1$\mu$m, e corrisponde ad un 
tempo di attraversamento $d/c$ = 3.6 fs.  La band-gap dovuta al multistrato
si estende per 
un'intervallo di lunghezze d'onda che va da 600 a 800 nm, con un minimo 
dell'$1 \%$ a 692 nm nella trasmissione. La faccia opposta 
\`e invece ricoperta di materiale antiriflessione. Spostando il supporto
\`e possibile introdurre, o no, una zona di evanescenza sul percorso 
del fotone. I due fotoni vengono poi fatti ri-incrociare, attraverso alcuni 
specchi, in modo da farli passare contemporaneamente per un divisore di fascio
al 50 $\%$. Sulle uscite del divisore di fascio sono posti due rivelatori, che 
registrano una coincidenza se i due fotoni arrivano entro 500 ps l'uno 
dall'altro. \par
Quando i due elettroni arrivano simultaneamente all'interno del divisore 
di fascio, l'interferenza tra i due ha un 
effetto distruttivo, annullando cos\'{\i} la coincidenza. 
Aumentando o diminuendo, attraverso un prisma mobile, 
il cammino di uno dei due elettroni \`e possibile stabilire quando 
i due attraversano contemporaneamente il divisore di fascio, o quale dei due 
arriva prima.\par 
Introdotto quindi lo spessore di multistrato sul cammino di uno dei due 
elettroni, \`e possibile misurarne, in questo modo, l'anticipo o il ritardo che 
questo ha rispetto all'altro.\\

\

Figura II-10.\\

Didascalia della Fig.II-10:\hfill\break
\{Profili dei conteggi di coincidenze con e senza la barriera  riflettente 
(rispettivamente curva pi\'u in alto e curva pi\'u in basso). Come si pu\`o 
vedere la curva pi\'u alta mostra un anticipo di circa {\rm 1.1 ($\pm$ 0.3)
fs} rispetto all'altra.\} \\

\

\h In Figura II.10 \`e riportato il risultato ottenuto dai tre autori sia nel caso 
di presenza della ``barriera", sia in sua assenza.\par
Si pu\`o dunque vedere che nel caso di presenza del multistrato, i fotoni che lo 
attraversano arrivano, in media, circa 1.1 fs prima degli altri. L'errore 
considerato dagli autori \`e di 0.3 fs. Ci\`o dimostra che  che il tempo 
impiegato da ogni singolo fotone ad attraversare la barriera \`e di circa 7 
deviazioni standard pi\'u piccolo del tempo che lo stesso elettrone 
impiegherebbe per attravesare lo spazio da questa occupato, ma in sua assenza, 
alla velocit\`a $c$. \par 
Gli autori fanno notare inoltre che, in questo caso, poich\'e per quasi tutto 
l'intervallo della band-gap il coefficiente di trasmissione varia molto 
lentamente, mantenendosi quasi costante al variare dell'energia, non \`e 
possibile giustificare tale evento come apparente e dovuto all'attraversamento, 
preferenzialmente, di fotoni pi\'u energetici. Inoltre notiamo che tutti i 
fotoni nell'aria viaggiano  circa alla stessa velocit\`a, molto prossima a 
quella della luce, indipendentemente dalla loro energia.\par
Gli stessi autori, per\`o, sostengono la tesi opposta, in un articolo 
divulgativo apparso su Scentific American nell'Agosto del 1993.\\

\

{\bf 
II.3) Misure del gruppo di Firenze.}\\ \rm

Pur operando anch'esso con microonde, il gruppo di Firenze 
riprende sia dal punto di vista dell'analisi teorica, che nella procedura 
sperimentale un'idea di Brillouin$[*41]$,  sviluppata in seguito da 
Stevens$[*42,*43]$.
Secondo questi infatti, in un mezzo dispersivo, va considerata come 
velocit\`a di propagazione di un segnale elettromagnetico la velocit\`a di 
propagazione del fronte d'onda molto netto di un segnale a scalino,
e non la sua velocit\`a di gruppo. \par
Notiamo che, anche se un segnale del 
genere pu\`o essere anch'esso soluzione dell'equazione di Helmholtz, non 
esiste  per\`o un equivalente quantistico. 
La propagazione di un segnale del genere \`e, poi, 
molto pi\'u dispersiva di quella di altri tipi di segnali, ed \`e accompagnata 
sempre dall'apparire, dopo un certo intervallo di tempo, di alcuni precursori,
formati dalle frequenze pi\'u alte.\\

\

Figura II-11.\\

Didascalia della Fig.II-11:\hfill\break
\{Ritardi temporali in funzione della frequenza per una barriera di 
15 cm di lunghezza. In figura sono pure riportate le curve teoriche di 
$\tau^s$, $\tau_\Trm^\varphi$ e $\tau_\Trm^\Brm$.\} \\

\

\h Tralasciamo dunque le analisi teoriche sull'argomento, ma prendiamo ugualmente 
in esame i loro dati sperimentali, anche perch\'e in essi 
ci troviamo un confronto 
diretto con alcuni dei tempi precedentemente definiti nel primo capitolo.\par 
La procedura 
usata assomigia abbastanza a quella del gruppo di
Enders e Nimtz nella prima delle 
loro misure riportate.
Le guide d'onda sono in questo 
caso rispettivamente: di banda $X$ quella esterna 
e di banda $P$ quella interna (per le dimensioni vedere ref. [*44]).
In una prima serie di misure, un segnale a scalino, nell'intervallo di frequenze
9.4, 9.7 GHz viene fornito da un klinstron, mentre il segnale trasmesso viene
mandato  ad un oscilloscopio ad alta risoluzione per la misura dei ritardi di 
fase e dell'attenuazione del segnale.\par
I risultati dei ritardi di fase sono  
corretti dagli autori mediante la sottrazione del ritardo corrispondente alla 
misura fatta senza la guida di banda $P$.\ppar
In Figura II.11 sono riportati i valori dei ritardi di fase misurati
in funzione delle frequenze, insieme  con le curve teoriche di
$\tau^s$, $\tau_\Trm^\varphi$ e $\tau_\Trm^\Brm$. Vediamo, che anche se nessuna delle 
curve riproduce esattamente  i dati sperimentali, quella del \phtz
sembra si avvicini abbastanza ai dati.\\

\

Figura II-12.\\

Didascalia della Fig.II-12:\hfill\break
\{Come per la  Figura II.11, ma per {\rm L = 10 cm}.\} \\

\

\h La Figura II.12 si riferisce, invece, al caso 
di  L = 10 cm. In questo caso per\`o
i punti sperimentali ricadono tra la curva relativa a $\tau_\Trm^\Brm$ e quella del 
\pht, mentre $\tau^\Srm$ rimane inadeguato. Gli autori attribuiscono tale 
comportamento alla giunzione tra le due guide, infatti, a quanto pare, non 
viene usata nessuna tecnica di calibrazione del sistema. L'effetto 
aumenterebbe, secondo gli autori, al diminuire della lunghezza della 
barriera.\\

\

{\bf
II.4) Conclusioni.}\\ \rm
Come abbiamo visto, il problema posto all'inizio, di quanto tempo impieghi una 
particella ad attraversare una zona classicamente proibita, bench\'e 
sostanzialmente risolto, rimane ancora controverso. \par
Se da un lato infatti esiste una evidenza sperimentale  di gran lunga a favore 
della definizione di \pht, dall'altro questa stessa definizione, e gli stessi 
risultati sperimentali, portano a dover accettare l'insorgere
di velocit\`a di gruppo maggiori di quella della luce nel vuoto, a causa 
di quello che Olkhovsky e Recami$[5]$ hanno chiamato ``effetto Hartman". \par 
Va comunque notato che, l'apparizione di velocit\`a di gruppo 
Superluminali o addirittura {\em negative}, nell'ottica classica, non \`e un 
fenomeno nuovo, e fu anche questo studiato ed osservato sperimentalmente
in lavori$[45*,47*]$ che solo di recente stanno richiamando l'attenzione che
meritavano.  Infatti, come sappiamo, in un mezzo dispersivo 
lineare, non assorbente: 
$$v_\grm=\displaystyle{\drm\omega\over\drm k}=
	\displaystyle{c\over n(\omega )+ \omega n^\prime (\omega)}\ ,$$
e, in regioni forte dispersione anomala (in prossimit\`a di risonanze), la 
velocit\`a di gruppo pu\`o eccedere $c$ o, come gi\`a detto, divenire negativa.
\`E comunque dimostrato, per\`o, che in questi casi la velocit\`a di gruppo 
perde di validit\`a fisica e nessun segnale pu\`o in effetti essere trasportato 
dal mezzo a velocit\`a maggiore di $c$ [si vedano anche il Cap.7 del 
{\em Classical Electrodynamics} del Jackson, o il Cap.3 del Sommerfeld]. \par
In tal caso, inoltre, il fenomeno \`e 
facilmente spiegato attraverso un reshaping dell'impulso, che subisce 
un'attenuazione delle componenti meno energetiche  e pi\'u lente. Il segnale 
uscente sembra quindi viaggiare  a velocit\`a  apparentemente maggiori 
di quella a cui hanno realmente viaggiato le sue componenti pi\'u energetiche. 
\ppar
Nel tunnelling, per\`o, abbiamo visto che (sia per le particelle, sia nel 
caso elettromagnetico) possiamo sempre metterci in condizioni da evitare 
effetti dovuti ad eventuali reshaping, o alla trasmisiione di particelle gi\`a
inizialmente pi\'u veloci.\par 
Ci\`o \`e anche confermato sperimentalmente dal fatto 
che, come abbiamo visto soprattutto nelle misure di Enders e Nimtz, la forma 
dei segnali rimane inalterata. \ppar
Notiamo per\`o che il processo di attraversamento della barriera, da parte di 
una particella, \`e comunque un 
processo statistico, nel senso che non possiamo mai sapere {\em a priori} 
quale particella attraversa la barriera. Inoltre, l'apparire di velocit\`a 
Superluminali  si ha proprio 
quando la probabilit\`a associata a tale evento \`e molto bassa. \par
Secondo alcuni autori, quindi, una giustificazione della violazione del 
principio di causalit\`a potrebbe essere che tale fenomeno, in quanto non
controllabile, non pu\`o essere usato per trasmettere alcuna informazione.\par
In quest'ottica si inserisce il tipo di giustificazione che portano 
Steinberg et al.$[10]$.  Secondo questi ultimi, nei processi di tunnelling in 
questione (quelli 
Superluminali), i relativi tempi di attraversamento andrebbero visti come 
{\em ``valori deboli"} della misura. Il concetto di {\em valore debole},
o {\em misura debole}, \`e stato introdotto nel 1988 da Aharonov, Albert e 
Vaidman$[49*,50*]$, partendo dalla teoria ``classica" della misura di 
von Neumann$[51*]$.\par
Secondo i due autori quando facciamo una misura, su di un sottoinsieme con 
bassa probabilit\`a associata (come quelo delle particelle trasmesse), e questo 
sottoinsieme proviene da un insieme  su cui \`e stata effettuata una {\em 
``misura debole"} (una misura cio\`e con una grossa inderminazione, che lasci 
lo stato del sistema quasi imperturbato), \`e possibile ottenere come risultato 
della misura sul sottoinsieme, un valore completamente diverso da tutti gli 
autovalori accessibili al sistema. Tale valore non sarebbe comunque un valore
realmente assunto dal sistema. \par
Secondo altri autori, invece, le velocit\`a Super-luminali 
legate al tunnelling sarebbero proprio reali e quello che va reinterpretato \`e 
il principio di causalit\`a.$[46*]$  A quest'ultimo proposito, ricordiamo che 
la Relativit\'a Speciale pu\`o essere estesa ---senza variarne i consueti
Postulati--- in modo da inglobare i moti Superluminali; la ``Relativit\`a 
Estesa"[47*], in altre parole, include i tachioni senza sostanzialmente violare
la Relativit\`a di Einstein, ma solo estendendola al nuovo dominio di
velocit\`a. In particolare, si possono risolvere le questioni causali.[48*]
\par
\`E anche utile ricordare che la stessa Relativit\`a Estesa prevede
per oggetti Superluminali, sulla base di semplici cosiderazioni 
geometrico-classiche, la transizione da velocit\`a di gruppo positive a 
velocit\`a negative (quando si ``oltrepassi" la situazione di
velocit\`a infinita): si vedano le refs.[47*].  Dato che a velocit\`a negative
corrispondono tempi di transito negativi[29*], ci\`o diviene interessante
alla luce di esperimenti meno recenti[42*] e pi\'u recenti[34-38]. \\

\

{\bf 
Ringraziamenti.}\\ \rm
Il presente lavoro si basa sulla Tesi di laurea di G.Privitera (``Tempi di
Tunnelling"; Universit\`a di Catania, 1995), avente come relatori E.Recami e 
G.D.Maccarrone. \ Siamo molto grati, per stimolanti e utili discussioni, 
e per la collaborazione
scientifica, a A.Agresti, M.Baldo, P.Barbero, R.Bonifacio, L.Bosi, 
G.Cavalleri, R.Chiao, G.Degli Antoni, F.Fontana, R.Garavaglia, A.Gigli
Berzolari, H.Hern\'andez, L.C.Kretly, J.-y.Lu, G.D.Maccarrone, G.Nimtz, 
A.Shaarawi, P.Saari, G.Salesi, A.M.Steinberg, M.T.Vasconselos e 
A.K.Zaichenko.\\

\

\centerline{{\bf Bibliografia}}

\

\noindent
[1]  E.U. Condon:  Rev. Mod. Phys. \ner 3, \rm 43, 
(1931)\parn       
[2]  L.A. MacColl:  Phys. Rev. \ner 40, \rm 621,
(1932)\parn
[3] V.S. Olkhovsky, E. Recami and A.I. Gerasimchuk: Nuovo Cimento A22 (1974)
263; \  E. Recami: ``A time operator and the time--energy uncertainty
relation", in {\em The Uncertainty Principle and Foundations of Quantum
Mechanics}, ed. by W.C.Price and S.S.Chissick (J.Wiley; London, 1977), p.21;
E. Recami: ``An operator for Time", in {\em Proceedings of the
Karpacz Winter School (Recent Developments in Relativistic QFT and its
Application, vol. 2)}, ed. by W.Karwowski (Wroclaw Univ. Press; Wroclaw,
1976), p. 251, and refs. therein; \ V.S. Olkhovsky: Sov. J. Part. Nucl. 15 
(1984) 130; \ Nukleonika 35 (1990) 99-144; \ ``The study of nuclear 
reactions by their temporal properties", in {\em Nuclear Reaction Mechanisms}, 
ed. by D.Seeliger and H.Kalka (World Scientific; Singapore, 1991), p.15.\parn
[4]  S. Collins, D. Lowe, J.R. Barker:  J. Phys. 
\ner C 20, \rm 6233, (1989); \  E.H. Hauge, J. A. Stovneng: 
Rev. Mod. Phys. \ner 61, \rm 917 (1989); \ A.P. Jauho: ``Tunneling times in
heterosctructures: A critical review", in {\em Hot Carriers in Semiconductor
Nanostructures: Physics and Applications} (A.T.T. Company; 1992) pp.121-150.
\parn
[5]  V.S. Olkhovsky, E. Recami:  Phys. Reports 
\ner 214, \rm 339 (1992)\parn
[6]  R. Landauer, Th. Martin:  Rev. Mod. Phys. 
\ner 66, \rm 217 (1994)\parn
[7]  R.P. Feynman, R.B. Leighton, M. Sands:  
{\em The Feynman lectures on Physics},  
(Addison-Wesley; 1977), vol.2, pp.24-27\parn
[8]  T.E. Hartman:  J. Appl. Phys. 
\ner 33, \rm 3427 (1962)\parn
[9]  J.R. Fletcher:  J. Phys.
\ner C 18, \rm 155 (1985)\parn
[10] See for instance  A.M. Steinberg:  J.
Physique-I \ner 4, \rm 1813 (1994), and refs. therein; \ Phys. Rev. A52
(1995) 32-52. \ Cf. also K. Hauss and P. Busch: Phys. Lett. A185 (1994)
9-13.\parn
[11]  E.H. Hauge, J.P. Falck, T.A. Fjeldly:
Phys. Rev. \ner B 36, \rm 4203 (1987)\parn
[12]  C.R. Leavens, G.C. Aers:  Phys. Rev.
\ner B 89, \rm 1202
(1989)\parn
[13]  Th. Martin, R. Landauer:  Phys. Rev. Lett.
\ner 49, \rm 1739 (1982)\\ \parn
[14]  M. B\"uttiker, R. Landauer:  Phys. Rev. Lett.
\ner 49, \rm 1739 (1982)\parn\
[15]  M. B\"uttiker, R. Landauer:  Phys. Scr.
\ner 32, \rm 429
(1985)\parn
[16]  M. B\"uttiker, R. Landauer:  I.B.M. J. Res.
Dev. \ner 30, \rm 451 (1986)\parn
[17]  A.J. Baz':  Sov. J. Nucl. Phy.
\ner 4, \rm 182 (1967)\parn
[18]  A.J. Baz':  Sov. J. Nucl. Phy.
\ner 5, \rm 161 (1967)\parn
[19]  V.E. Rybachenko:  Sov. J. Nucl. Phy.
\ner 5, \rm 635 (1967)\parn
[20]  J.J. Sakurai:
{\em Meccanica Quantistica Moderna}, 
(Zanichelli, 1990) \rm pp.75-77 \parn
[21]  M. B\"uttiker:  Phys. Rev. \ner B 27, 
\rm 6178 (1983)\parn
[22]  J.P. Falck, E.H. Hauge:  Phys. Rev. 
\ner B 38, \rm 3287 (1988)\parn
[23]  D. Sokolovski, L.M. Baskin:  Phys. Rev. 
\ner A 36, \rm 4604 
(1987)\parn
[24]  P. H\"anggi:  In {\em Lectures on  Path Integration} 
(World Scientific; London, 1991), \rm pp.352 \parn
[25]  P. Sokolovski, J.N.L. Connor:  Phys. Rev. 
\ner A47, \rm 4667 (1993). \ See also H.A.Fertig: Phys. Rev. Lett. 65 (1990)
2321; Phys. Rev. B47 (1993) 1346.\parn
[26]  C.R. Leavens, G.C. Aers:  In
{\em Scanning Tunneling Microscopy -- III}, ed. by
R. Wiesedanger, H.J. G\"untherodt (Springer;
New York, 1993) pp. 105 \parn
[27]  F.T. Smith:  Phys. Rev. \ner 118,
\rm 349 (1960)\parn
[28]  V.S. Olkhovsky, E. Recami, A.K. Zaichenko:  
Sol. State Com.  \ner 89, \rm 31 (1994)\parn
[29]  V.S. Olkovsky, R. Recami, F. Raciti, A.K. Zaichenko: 
 J. Phys. \ner 5, \rm 1351 (1995)\parn
[30]  W. Jaworski, D.M. Wordlawd:  Phys. Rev. 
\ner A 37, \rm 2834 (1998)\parn
[31]  C.R. Leavens:  Solid State Com. 
\ner 85, \rm 115 (1993)\parn
[32]  R.S. Dumont, T.L. Marchioro:  
Phys. Rev. \ner A 47, \rm 85 (1993)\parn
[33]  A.M. Steinberg, P.G. Kwiat, R.Y. Chiao:
Phys. Rev. Lett.  \ner 71, \rm 708 (1993)\parn
[34] A.Enders and G.Nimtz:  J. de
Physique-I  2 (1992) 1693; 3 (1993) 1089;
 Phys. Rev.  B47 (1993) 9605;
 Phys. Rev.  E48 (1993) 632; G.Nimtz,
A.Enders and H.Spieker:  J. de
Physique-I 4 (1994) 1; W.Heitmann and G.Nimtz:
Phys. Lett. A196 (1994) 154; G.Nimtz,
A.Enders and H.Spieker: in  {\em Wave and Particle in
Light and Matter}  (Proceedings of the Trani Workshop,
Italy, Sept.1992), eb. by A.van der Merwe and A.Garuccio (Plenum; New York,
in press); H.Aichmann and G.Nimtz, ``Tunnelling of a
FM-Signal: Mozart 40," submitted for pub.; G.Nimtz and W.Heitmann:
Prog. Quant. Electr. 21 (1977) 81-108.\parn
[35] A.M.Steinberg, P.G.Kwiat and R.Y.Chiao: ref.[33]; \ R.Y.Chiao,
P.G.Kwiat and A.M.Steinberger:  Scientific American
269 (1993), issue no.2, p.38; \  A.M.Steinberg and R.Y.Chiao: Phys. Rev.
A51 (1995) 3525-3528. \ Cf. also P.G.Kwiat, A.M.Steinberg, R.Y.Chiao,
P.H.Eberhard and M.D. Petroff: Phys. Rev. A{\bf 48} (1993) R867; \ E.L.Bolda,
R.Y. Chiao and J.C. Garrison: Phys. Rev. A{\bf 48} (1993) 3890;
A.M.Steinberg: Phys. Rev. Lett. 74 (1995) 2405; \ R.Y.Chiao and
A.M.Steinberg: ``Tunneling times and superluminality", in {\em Progress
in Optics}, vol.37 (1997), ed. by E.Wolf.\parn
[36] A.Ranfagni, P.Fabeni, G.P.Pazzi and D.Mugnai:
Phys. Rev.  E48 (1993) 1453.\parn
[37] Ch.Spielmann, R.Szipocs, A.Stingl and F.Krausz:  Phys.
Rev. Lett.  73 (1994) 2308.\parn
[38]  Scientific American: an article
in the Aug. 1993 issue;  Nature:
comment ``Light faster than light?" by R.Landauer, Oct. 21,
1993;  New Scientist:  editorial
``Faster than Einstein" at p.3, plus an article by J.Brown at
p.26, April 1995.\parn
[39] Th. Martin and R. Landauer: Phys
Rev.  45A (1992) 2611; R.Y.Chiao, P.G.Kwiat and
A.M. Steinberg:  Physica  B175 (1991)
257; A. Ranfagni, D. Mugnai, P. Fabeni and G.P. Pazzi:
Appl. Phys. Lett.  58 (1991) 774; and refs. therein. \ See also
A.M.Steinberg: {\em Phys. Rev.} A52 (1995) 32.\parn
[40] A. Agresti, V.S. Olkhovsky and E. Recami: (to be submitted for pub.)\parn
[41] H.M. Brodowsky, W. Heitmann and G. Nimtz: ``Comparison of experimental
microwave tunnelling data with calculations based on Maxwell equations",
preprint (University of Cologne; 20 Aug.1996) submitted to Elsevier
Preprint.\parn
[42] A.P.L. Barbero, H. Hern\'andez F., and E.Recami: (to be submitted
for pub.).

\

[*39]  H.J. Eul, B. Schick:  IEEE-MTT \ner 39,
\rm 724 (1991)\parn
[*40]  L. Brillouin: {\em Waves Propagation and Group
Velocity} (Academic Press; New York, 1960)\parn
[*41]  K.W. H.Stevens: Europ. J. Phys.
\ner 1, \rm 98 (1980)\parn
[*42]  K.W.H. Stevens:
J. Phys. \ner C 16, \rm 3649 (1983)\parn
[*43]  A. Ranfagni, D. Mugnai, P. Fabeni, G.P. Pazzi, G. Naletto,
C. Sozzi: Physica \ner B175, \rm 283 (1991)\parn
[*44]  C.G.B. Garret, D.E. McCumber:
Phys. Rev. \ner A1, \rm 305 (1970)\parn
[*45]  S. Chu, S. Wong:
Phys. Rev. Lett. \ner 49, \rm 1293 (1982)\parn
[*46]  E.L. Bolda, R.Y. Chiao:
Phys. Rev. \ner A 48, \rm 3890 (1993)\parn
[*47]  A.M. Steinberg: \ner , \rm \parn
[*48]  A.M. Steinberg: \ner , \rm \parn
[*49]  Y.Aharonov,D.Z.Albert and L.Vaidman; Phys. Rev. Lett. 60 (1988)
1351; \ Y. Aharonov, L. Vaidman:  Phys. Rev. \ner A 41, \rm 11 (1990)\parn
[*50]  I.M. Duck, P.M. Stevenson, E.C.G. Sudarshan:
Phys. Rev. \ner D 40, \rm 40 (1989)\parn
[*51]  J. von Neumann:
{\em Mathematical Fondation of Quantum Meccanics},
(Princeton Univ. Press, Princeton, 1983)\parn
[*52]  R. Landauer:
Nature \ner 341, \rm 567 (1989) \parn
[*53]  Ch. Spieldmann et al.:
Phis. Rev. Lett. \ner 73 , \rm 2308 \parn
[*54]  A. Ranfagni, D. Mugnai, A. Agresti:
Phis. Lett. \ner A 158, \rm 161 (1991)\parn
[*55]  D. Mugnai, A. Ranfagni, R. Ruggeri, A. Agresti, E. Recami:
 Phis. Lett. \ner A 209, \rm 227 (1995)\parn

\

[41*]  C.G.B. Garret, D.E. McCumber:
Phys. Rev. \ner A1, \rm 305 (1970)\parn
[42*]  S. Chu, S. Wong:
Phys. Rev. Lett. \ner 49, \rm 1293 (1982)\parn
[43*] Y.Aharonov,D.Z.Albert and L.Vaidman; Phys. Rev. Lett. 60 (1988)
1351; \ Y. Aharonov, L. Vaidman:  Phys. Rev. \ner A 41, \rm 11 (1990)\parn
[44*]  I.M. Duck, P.M. Stevenson, E.C.G. Sudarshan:
Phys. Rev. \ner D 40, \rm 40 (1989)\parn
[45*]  J. von Neumann:
{\em Mathematical Fondation of Quantum Meccanics}
(Princeton Univ. Press; Princeton, 1983)\parn
[46*]  D. Mugnai, A. Ranfagni, R. Ruggeri, A. Agresti, E. Recami:
Phis. Lett. \ner A 209, \rm 227 (1995)\parn
[47*] See, e.g., E. Recami: ``Classical tachyons and possible
applications,"  Rivista Nuovo Cim. 9 (1986), issue no.6, pp.1-178, and
refs. therein; \ E. Recami and W.A. Rodrigues Jr.: ``A model theory for
tachyons in two dimensions", in  {\em Gravitational Radiation and
Relativity,} ed. by J.Weber and T.M.Karade (World
Scient.; Singapore, 1985), pp.151-203.\parn
[48*] E.Recami: ``Tachyon kinematics and causality",
Foundation of Physics 17 (1987) 239-296; \ ``The Tolman `Anti-telephone'
paradox: Its solution by tachyon mechanics,"  Lett. Nuovo Cimento 44 (1985)
587-593.\parn

%%%Wimmel
%%%Ranfagni e Mugnai: Phys. Rev. E52 (1995) 1128.
\end{document}